\documentclass[11pt]{revtex4}
\usepackage{amsmath}
\usepackage{amsfonts}
\usepackage{amssymb}
\usepackage{amsxtra}
\usepackage{supertabular}
\usepackage[pdftex]{graphicx}
\usepackage{tikz}
\usetikzlibrary{arrows,snakes}
\usepackage{subfig}
\usepackage{amsxtra} 
\begin{document}
\title{Understanding the second quantization of fermions in Clifford and in Grassmann space\\
New way of second quantization of fermions\\
Part I}
\author{N.S. Manko\v c Bor\v stnik${}^1$ and H.B.F. Nielsen${}^2$
\\
${}^1$
University of Ljubljana, Slovenia\\
${}^2$Niels Bohr Institute, Denmark
}
%

\begin{abstract}
Both algebras, Clifford and Grassmann,  offer ''basis vectors'' for 
describing the internal degrees of freedom of  fermions~\cite{norma92,%
norma93,nh2018}. The oddness of the ''basis vectors'', transfered to 
the creation operators, which are tensor products of the finite number of 
''basis vectors'' and the infinite number of momentum basis, and  to their 
Hermitian conjugated partners annihilation operators, offers 
the second quantization of fermions without postulating the conditions 
proposed by Dirac~\cite{Dirac,BetheJackiw,Weinberg}, enabling
the explanation of the Dirac's postulates. 
But while the Clifford fermions manifest the half integer spins --- in 
agreement with  the observed properties of quarks and leptons and 
antiquarks and antileptons --- the ''Grassmann fermions" manifest 
the integer spins.
In Part I properties of the creation and 
annihilation operators of integer spins "Grassmann fermions" are 
presented and the proposed equations of motion solved. The anticommutation
relations of second quantized integer spin fermions are shown  when applying
on the vacuum state as well as when applying on the Hilbert space 
of the infinite number of "Slater determinants" with all 
the possibilities of empty and occupied "fermion states".  
In Part II the conditions are discussed under which the Clifford algebras 
offer the appearance  of the second quantized fermions, enabling as well the 
appearance of families. In both parts, Part I and Part II, the relation between 
the Dirac way and our way of the second quantization of fermions is presented.
%
\keywords{Second quantization of fermion fields in Clifford and in
  Grassmann space \and
Spinor representations in Clifford and in Grassmann space \and
Explanation of the Dirac postulates \and 
Kaluza-Klein-like theories \and 
Higher dimensional spaces \and Beyond the standard model}
\end{abstract}

\keywords{Second quantization of fermion fields in Clifford and in
  Grassmann space,
Spinor representations in Clifford and in Grassmann space,
Explanation of the Dirac postulates,
Kaluza-Klein-like theories,
Higher dimensional spaces,
Beyond the standard model}

\maketitle

\section{Introduction}
\label{introduction}


In a long series of works we, mainly one of us N.S.M.B.~(\cite{norma92,%
norma93,IARD2016,n2014matterantimatter,nd2017,n2012scalars,%
JMP2013,normaJMP2015,nh2018} and the references therein), have found 
phenomenological success with the model named by N.S.M.B the 
{\it spin-charge-family}  theory,  with fermions, the internal degrees of 
freedom of which is describable with the Clifford algebra of all linear 
combinations of products of $\gamma^a$'s in $d=(13+1)$ (may be  with 
d infinity), interacting with only gravity. The spins of fermions from higher 
dimensions, $d>(3+1)$, manifest in $d=(3+1)$ as charges of the 
{\it standard model}, gravity in higher dimensions manifest as the {\it standard 
model} gauge vector fields as well as the scalar Higgs  and Yukawa couplings.

There are two anticommuting kinds of algebras, the Grassmann algebra and 
the Clifford algebra (of two independent subalgebras), expressible with each 
other. The Grassmann algebra, with elements $\theta^a$, and their 
Hermitian conjugated partners $\frac{\partial}{\partial \theta^a}$~\cite{nh2018}, 
can  be used to describe the internal space of fermions with the integer spins 
and charges in the adjoint representations, the two Clifford algebras, we call 
their elements $\gamma^a$ and $\tilde{\gamma}^a$, can each of them be 
used to describe half integer spins and charges in fundamental representations. 
The Grassmann algebra  is equivalent to the two Clifford algebras and opposite.

The two papers explain how do the oddness of the internal space of fermions 
manifests in the single particle   wave functions, relating the oddness of the 
wave functions to the corresponding creation and annihilation operators of the 
second quantized fermions, in the Grassmann case and in the Clifford case,
explaining therefore the postulates of Dirac for the  second quantized fermions.
We also show that the requirement that the Clifford odd algebra represents
the observed quarks and leptons and antiquarks and antileptons reduces the
Clifford algebra for the factor of two, reducing at the same time the
Grassmann algebra, disabling the possibility for the integer spin fermions.

In this paper  it is demonstrated how do the Grassmann algebra --- in Part I --- and 
the two kinds of the Clifford algebras --- in Part II --- if used to describe the internal 
degrees of freedom of fermions, 
take care of the second quantization of fermions without postulating  anticommutation 
relations~\cite{Dirac,BetheJackiw,Weinberg}. 
Either the odd Grassmann algebra or the odd Clifford algebra offer namely  the appearance 
of the creation operators, defined on the tensor products of the ''basis vectors'' of the 
internal space and of the momentum space basis. These creation operators, together with 
their Hermitian conjugated partners anihilation operators, inherit oddness  from the ''basis 
vectors'' determined by the odd Grassmann or the odd Clifford algebras, fulfilling 
correspondingly, the anticommutation relations postulated by Dirac for the second 
quantized fermions, if they apply on the 
corresponding vacuum state, Eq.~(\ref{vactheta}) (defined by the sum of products of all the annihilation  times the corresponding Hermitian conjugated creation operators).
Oddness of the ''basis vectors'', describing the internal space of fermions, guarantees the 
oddness of all the objects entering  the tensor product.

In $d$-dimensional Grassmann space of anticommuting coordinates $\theta^{a}$'s,
$i=(0,1,2,3,5,\cdots,d)$, there are $2^d$ "basis vectors", which are superposition of 
products of $\theta^{a}$. One can arrange them into the odd and the even  irreducible representations with respect to the Lorentz group.  
There are as well derivatives with respect to $\theta^{a}$'s, $\frac{\partial}{\partial \theta_{a}}$'s,
taken in Ref.~\cite{nh2018} as, up to a sign, Hermitian conjugated to $\theta^{a}$'s, 
$(\theta^{a \dagger} = \eta^{a a} \frac{\partial}{\partial \theta_{a}}$,  
$\eta^{a b}=diag\{1,-1,-1,\cdots,-1\}$), which form again $2^d$ "basis vectors". 
Again half of them odd and half of them even (the odd Hermitian conjugated 
to odd products of $\theta^a$'s, the even Hermitian conjugated to the even 
products of $\theta^a$'s). 
Grassmann space offers correspondingly $2\cdot 2^d$ degrees of freedom. 

There are two kinds of the Clifford "basis vectors", which are expressible with 
$\theta^{a}$
and $\frac{\partial}{\partial \theta_{a}}$: $\gamma^{a}= (\theta^{a} + 
\frac{\partial}{\partial \theta_{a}})$, $\tilde{\gamma}^{a}= i \,(\theta^{a} -  
\frac{\partial}{\partial \theta_{a}})$~\cite{norma93,nh02,nh03}. They are, up to 
$\eta^{aa}$, Hermitian operators.  
Each of these two kinds of the Clifford algebra objects has $2^d$ operators. "Basis 
vectors" of Clifford algebra have together again $2\cdot2^d$ degrees of freedom.
 
%
There is the {\it odd algebra} in all  three cases,  $\theta^{a}$'s,  
$\gamma^{a}$'s, $\tilde{\gamma}^{a}$'s, which  if used to generate the creation 
and annihilation operators for fermions, and correspondingly the single fermion states,
leads to the Hilbert space of  second quantized 
fermions obeying the anticommutation relations of Dirac~\cite{Dirac} without 
postulating these relations: the anticommutation properties follow from the oddness
of the ''basis vectors'' in any of these algebras.

Let us present steps which lead to the second quantized fermions:\\ 
{\bf i.}  The internal space of a fermion is described by either Clifford or Grassmann 
algebra of an odd Clifford character (superposition of an odd number of Clifford "coordinates" 
(operators) $\gamma^{a}$'s or of an odd number of Clifford "coordinates" (operators) 
$\tilde{\gamma}^{a}$'s)
or of an odd Grassmann character (superposition of an odd number of Grassmann 
"coordinates" (operators) $\theta^a$'s).\\
{\bf ii.} The eigenvectors of all the (chosen) Cartan subalgebra members of the 
corresponding Lorentz algebra 
are used to define the "basis vectors" in the odd part of  internal space of  fermions. 
(The Cartan subalgebra is in all three cases chosen in the way to be in agreement with the ordinary choice.)
The algebraic application of this "basis vectors" on the corresponding vacuum state (either
 Clifford $|\psi_{oc}>$, defined in Eq.~(18) of Part II, or Grassmann $|\phi_{og}>$, 
Eq.~(\ref{vactheta}), which  is in the Grassmann case 
just the identity) generates the "basis states",  describing
the internal degrees of freedom of fermions. The members of the "basis vectors" manifest 
together with their Hermitian conjugated partners properties of creation and annihilation operators  which anticommute, Eq.~(\ref{ijthetaprod}) in Part I and Eq.~(18) in Part II, when
applying on the corresponding vacuum state, due to the algebraic properties of the 
odd products of the algebra elements.\\
{\bf iii.} The plane wave solutions of the corresponding  Weyl equations (either Clifford,  Eq.~(23) or Grassmann, Eq.~(\ref{Weylgrass})) for free massless fermions are the 
tensor products of the superposition of the members of the "basis vectors" and of the 
momentum basis. The coefficients of the superposition correspondingly depend on a chosen momentum $\vec{p}$, with $|p^0| =|\vec{p}|$, for any of continuous
many moments $\vec{p}$.\\
{\bf iv.} The creation operators defined on the tensor products, $*_{T}$, of superposition of
finite number of  ''basis vectors'' defining the final internal space and of the infinite (continuous) momentum space, Eq.~(24) in the Clifford case and  Eq.~(\ref{ptheta}) in the 
Grassmann case, have infinite basis.  \\ 
{\bf v.} Applied on the vacuum state these creation operators form anticommuting single fermion states of an odd Clifford/Grassmann character. \\
{\bf vi.} The second quantized Hilbert space ${\cal H}$ consists of  
"Slater determinants" with no single particle state occupied (with no creation operators 
applying on the vacuum state), with one single particle state occupied (with one creation operator applying on the vacuum state), with two single particle states occupied  (with two creation operator applying on the vacuum state), and so on. ''Slater determinants'' can 
as well be represented as the tensor product multiplication of all possible single
particle states of any number.\\
{\bf vii.} The creation operators together with their Hermitian conjugated partners annihilation 
operators fulfill, due to the oddness of the ''basis vectors'', while the momentum
part commutes, the anticommutation relations, 
postulated by Dirac for second quantized fermion fields, not only when they apply on the
vacuum state, but also when they apply on the Hilbert space ${\cal H}$,  
Eq.~(39) in the Clifford case and Eq.~(\ref{ijthetaprodgenHT}) in the Grassmann case.
%
In the Clifford case this happens only after ''freezing out'' half of the Clifford  space, 
as it is shown in Part II, Sect. 2.2, what brings besides the correct anticommutation 
relations also  the ''family'' quantum number to each irreducible representation of 
the Lorentz group of the remaining internal space. \\
The oddness of the creation operators forming the single fermion states of an odd 
character, transfers to the application of these creation operators on the Hilbert 
space of the second quantized fermions in the Clifford and in the Grassmann case.\\
 %
{\bf viii.} Correspondingly the creation and annihilation operators with the internal 
space described by either odd Clifford or odd Grassmann algebra, since fulfilling 
the anticommutation relations required for the second quantized fermions without 
postulates, explain the Dirac's postulates for the second quantized fermions.\\
%




 
 In the subsection~\ref{HNsubsection} of this section we discuss in a generalized 
way our assumption, that the 
oddness of the ''basis vectors'' in the internal space transfer to the corresponding
creation and annihilation operators determining the second quantized single
fermion states and correspondingly the Hilbert space of the second quantized 
fermions.
 
We present in Sect.~\ref{propertiesGrass0} properties of the Grassmann odd (as well as,  
for our study of anticommuting ''Grassmann fermions'' not important, the Grassmann 
even) algebra and of the chosen "basis vectors" for even ($d=(2(2n+1), 
4n$), n is an integer) dimensional space-time, 
$d=(d-1)+1$, and illustrate anticommuting "basis vectors" on the case of $d=(5+1)$, 
Subsect.~\ref{propertiesGrass1}, chapter {\it A.b.}.

We define the action for the integer spin "Grassmann fermions" in 
Subsect.~\ref{actionGrass}. Solutions of the corresponding equations of motion, which
are the tensor products of finite number of ''basis vectors'' and of
infinite number of basis in momentum space,  define the creation operators depending on
internal quantum numbers and on $\vec{p}$ in $d$-dimensional space-time. We illustrate the corresponding superposition
of ''basis vectors'', solving the equation of motion in $d=(5+1)$ in chapter {\it B.a.}. 

We present in Sect.~\ref{Hilbertspace0} the Hilbert space ${\cal H}$ of the   
tensor multiplication of one fermion creation operators of all possible single particle 
states of an odd character and of any number,
representing ''Slater determinants''  with no "Grassmann fermion'' state occupied with "Grassmann fermions", with one "Grassmann fermion'' state occupied, with two 
"Grassmann fermion'' states occupied, up to the "Slater determinant" with all possible "Grassmann fermion'' states of each of infinite number of momentum  $\vec{p}$ occupied. 
The  Hilbert space ${\cal H}$ is the tensor product $\prod_{\infty}\otimes_{N}$
of finite number of ${\cal H}_{\vec{p}}$ of  a particular momentum $\vec{p}$, for 
(continues) infinite possibilities for $\vec{p}$. 
 
On ${\cal H}$ the creation and annihilation operators 
manifest the 
anticommutation relations of second quantized  "fermions" without any postulates.
These second quantized "fermion" fields, manifesting in the Grassmann case an 
integer spin, offer in $d$-dimensional space, $d > (3+1)$, the description of the 
corresponding charges  in adjoint representations. We follow in this paper to some 
extent  Ref.~\cite{nh2018}. 

In Subsect.~\ref{comparisonND} relation between the by Dirac postulated creation
 and annihilation operators and the creation and annihilation operators presented in 
 this Part I --- for integer spins ''Grassmann fermions'' --- are discussed.
 
 In Sect.~\ref{conclusionsGrass} we comment on what we have learned from the 
 second quantized "Grassmann fermion" fields with integer spin when internal degrees of freedom are described with Grassmann algebra and compare these recognitions with the recognitions, which the Clifford algebra is 
offering, discussions on which appear in Part II.

In Part II we present in equivalent sections 
properties of the two kinds of the Clifford algebras and discuss conditions under which 
odd products of odd elements (operators), $\gamma^a$ and $\tilde{\gamma}^a$'s  
of  the two Clifford algebras,  demonstrate the anticommutation relations
 required for the second quantized fermion fields on the Hilbert space ${\cal H} =
\prod_{\infty}\otimes_{N} {\cal H}_{\vec{p}}$, this time with the half integer spin, 
offering in $d$-dimensional space,
$d > (3+1)$, the description of charges, as well as the appearance of families of 
fermions~\cite{nh2018}, both needed to describe the properties of the observed quarks 
and leptons and antiquarks and antileptons, appearing in families.  

In Part II we discuss relations between the Dirac way of second quantization 
with postulates and our way using Clifford algebra.

This paper is a part of the project named the {\it spin-charge-family} theory of one 
of the authors (N.S.M.B.), so far offering the explanation for all the assumptions of 
the {\it standard model}, with the appearance of the scalar fields included.



%

The Clifford algebra offers in even $d$-dimensional spaces, $d \ge (13 +1)$ indeed, the description of the internal degrees of freedom for the second quantized fermions with the half integer spins, explaining all the assumptions of the {\it standard model}: The appearance of charges of the observed quarks and leptons and their families, as well as the appearance of the corresponding gauge fields, the scalar fields, explaining the Higgs scalar and the Yukawa couplings, and in addition the appearance of the dark matter, of the matter/antimatter asymmetry, offering several predictions~\cite{norma92,norma93,%
IARD2016,n2014matterantimatter,nd2017,n2012scalars,JMP2013,normaJMP2015,nh2017}. 
\subsection{Our main assumption and definitions}
\label{HNsubsection}

In this subsection we clarify how does the main assumption of Part I and 
Part II, {\it the decision to describe the internal space of fermions with 
the ''basis vectors'' expressed with the superposition of odd products of 
the anticommuting members of the algebra}, either the Clifford one or 
the Grassmann one, acting algebraically, $*_{A}$, on the internal 
vacuum state  $|\psi_{o}>$, relate to the creation and annihilation 
anticommuting operators of the second quantized  fermion fields. 

To appreciate the need for this kind of assumption, 
let us first have in mind that algebra with the product $*_A$ is 
only present in our work, usually not in other works, and thus has no 
well known physical meaning. It is at first a product by which you can 
multiply two internal wave functions $B_{i}$ and $B_{j}$ with each other,
\begin{eqnarray}
\label{HNA}
C_k&=& B_i *_{A}B_j\,,\nonumber\\
B_i *_{A}B_j&=& \mp B_j *_{A} B_i \,, \nonumber 
\end{eqnarray} 
the sign $\mp$ depends on whether $B_i$ and $B_j$ are products of 
odd or even number of algebra elements: The sign is $-$ if both 
are (superposition of) odd products of algebra elements, in all other 
cases the sign is $+$.

Let ${\bf R}^{d-1}$ define the external spatial or momentum space. 
Then the tensor product $*_{T}$ extends the internal wave functions 
into the wave functions ${\bf C}_{\vec{p} ,\,i}$ defined in both spaces  
\begin{eqnarray}
\label{HNT}
{\bf C}_{\vec{p}, \, i} =|\vec{p}>*_{T}|B_{i}>\,,\nonumber
\end{eqnarray}
where again $B_i $ represent the superposition of products of elements 
of the  anticommuting algebras, in our case either 
$\theta^a$ or $\gamma^a$ or $\tilde{\gamma}^a$, 
used in this paper.

We can make a choice of the orthogonal and normalized basis so that 
$<{\bf C}_{\vec{p}, i} | {\bf C}_{\vec{p'}, j} > = 
\delta(\vec{p} - \vec{p'}) \,\delta_{ij}$. Let us point out that either 
$B_{i}$ or ${\bf C}_{\vec{p}, \, i}$ apply algebraically  on the vacuum 
state, $B_i*_{A} |\psi_o>$ and ${\bf C}_{\vec{p}, \, i}*_{A}|\psi_o>$.

Usually a product of single particle wave functions is not taken to have any 
physical meaning in as far as most physicists simply do not work with such 
products at all. 

To give to  the algebraic product, $*_A$,  and to the tensor product, $*_{T}$, 
defined on the space of single particle wave functions, the physical meaning, 
we postulate the connection  between the anticommuting/commuting properties 
of the ''basis  vectors'', expressed with  the odd/even products of the anticommuting 
algebra elements and the corresponding creation operators, creating second 
quantized single fermion/boson states
\begin{eqnarray}
\label{HNb}
{\hat b}^{\dagger}_{{\bf C}_{\vec{p}, i}} *_{A}|\psi_o> &=& |\psi_{\vec{p}, i}>\,,\nonumber\\
{\hat b}^{\dagger}_{{\bf C}_{\vec{p}, i}}  *_{T}\,|\psi_{\vec{p'}, j}> &=& 0\,,\nonumber\\
{\rm if \,} \vec{p} &=& \vec{p'}\, {\rm and} \, i=j\,,\nonumber\\
{\rm in\; all\;other\;cases\;} & &{\rm it\;follows\,}\nonumber\\
{\hat b}^{\dagger}_{{\bf C}_{\vec{p}, i}}  *_{T}\, 
{\hat b}^{\dagger}_{{\bf C}_{\vec{p'}, j}} *_{A}|\psi_o>
&=& \mp \,{\hat b}^{\dagger}_{{\bf C}_{\vec{p'}, j}}  *_{T} \, 
{\hat b}^{\dagger}_{{\bf C}_{\vec{p}, i}}  *_{A}|\psi_o>\,,\nonumber
\end{eqnarray} 
with the sign $\pm $ depending on whether  ${\hat b}^{\dagger}_{{\bf C}_{\vec{p}, i}}$
have both an odd character, the sign is $-$, or not, then the sign is $+$.

To each creation operator ${\hat b}^{\dagger}_{{\bf C}_{\vec{p}, i}}$ its Hermitian 
conjugated partner  represents the annihilation operator ${\hat b}_{{\bf C}_{\vec{p}, i}}$
\begin{eqnarray} 
\label{HNC}
{\hat b}_{{\bf C}_{\vec{p}, i}} &=&({\hat b}^{\dagger}_{{\bf C}_{\vec{p}, i}})^{\dagger}\,,
\nonumber\\
{\rm with \; the}&&{\rm property}\nonumber\\
{\hat b}_{{\bf C}_{\vec{p}, i}}\, *_{A}\,|\psi_o> &=&0\,,\nonumber\\
{\rm defining\; the\; } && {\rm vacuum \; state\; as\;}\nonumber\\
|\psi_o> :&=&\sum_i \,(B_{i})^{\dagger}\, *_{A} \,B_{i}
\,|\;I>\nonumber
\end{eqnarray}
where summation $i$ runs over all different products of annihilation operator 
$\times$ its Hermitian conjugated  creation operator, no matter for what 
$\vec{p}$ , and $|\;I>$ represents the identity, $(B_{i})^{\dagger}$ represents
the Hermitian conjugated wave function to $B_i$. 

Let the tensor multiplication $*_{T}$ denotes also  the multiplication of any number
of single particle states, and correspondingly of any number of creation operators.

What further means that to each single particle wave function we define 
the creation operator ${\hat b}^{\dagger}_{{\bf C}_{\vec{p}, i}} $, applying 
in a tensor product from the left hand side on the second quantized 
Hilbert space --- consisting of all possible products 
of any number of the single particle wave functions --- adding to the Hilbert 
space the single particle wave function created by this particular creation 
operator. In the case of the second quantized fermions, if this particular 
wave function with the quantum numbers and $\vec{p}$  of 
${\hat b}^{\dagger}_{{\bf C}_{\vec{p}, i}} $ is already 
among the single fermion wave functions of a particular product of 
fermion wave functions, the action of the creation operator gives zero, 
otherwise the number of the fermion wave functions increases for one. 
In the boson case the number of boson second quantized wave functions 
increases always  for one. 

If we apply with the annihilation operator ${\hat b}_{{\bf C}_{\vec{p}, i}} $ 
on the second quantized Hilbert space, then the application gives a 
nonzero contribution only if  the particular products of the single particle 
wave functions do include the wave function with the quantum number 
$i$ and $\vec{p}$. 

In a Slater determinant formalism the single particle wave functions define
the empty or occupied places of any of infinite numbers of Slater determinants. 
The creation operator ${\hat b}^{\dagger}_{{\bf C}_{\vec{p}, i}} $ applies
on a particular Slater determinant from the left hand side. Jumping over 
occupied states to the place with its $i$ and $\vec{p}$. If this state is 
occupied, the application gives in the fermion case zero, in the boson case 
the number of particles 
increase for one. The particular Slater determinant changes sign in the 
fermion case if 
${\hat b}^{\dagger}_{{\bf C}_{\vec{p}, i}}$ jumps over odd numbers of occupied 
states.  In the boson case the sign of the Slater determinant does not change. 

When annihilation operator ${\hat b}_{{\bf C}_{\vec{p}, i}} $ applies on 
particular Slater determinant, it is jumping over occupied states to its own place.
giving zero, if this space is empty and decreasing the number of occupied states 
of this space is occupied. The Slater determinant changes sign in the fermion 
case, if the number of occupied states before its own space is odd. In the boson
case the sign does not change.

Let us stress that choosing antisymmetry or symmetry is a choice which we make 
when treating  fermions or bosons, respectively, namely 
the choice of using oddness or evenness of basis vectors, that is the choice of using 
odd products or even products  of algebra anticummuting elements. 

To describe the second quantized fermion states  we make a choice of the basis 
vectors, which are the superposition of the odd numbers of algebra elements,
of both Clifford and Grassmann algebras.

The creation operators and their Hermitian conjugation partners annihilation 
operators therefore in the fermion case anticommute. The single fermion 
states, which are the application of the creation operators on the vacuum 
state $|\psi_o>$, manifest correspondingly as well the oddness. 
The vacuum state, defined as the sum over all different products of 
annihilation $\times$ the corresponding creation operators, have an even
character.

Let us end up with the recognition:\\ 
One usually means antisymmetry when talking about Slater-\underline{determinants}
because otherwise one would not get determinants.

In the present paper~\cite{norma92,norma93,IARD2016,nh02} the choice of 
the symmetrizing versus antisymmetrizing  relates indeed the commutation versus anticommutation with respect to the a priori completely different product $*_A$, 
of anticommuting members of the Clifford or Grassmann algebra. The oddness or 
evenness  of these products transfer to quantities to which these algebras
extend.

%

\section{Properties of Grassmann algebra in even dimensional spaces}
\label{propertiesGrass0}

In Grassmann $d$-dimensional space there are $d$ anticommuting operators $\theta^{a}$, 
$\{\theta^{a}, \theta^{b}\}_{+}=0$, $a=(0,1,2,3,5,..,d)$, and $d$ anticommuting derivatives 
with respect to $\theta^{a}$, $\frac{\partial}{\partial \theta_{a}}$, 
$\{\frac{\partial}{\partial \theta_{a}}, \frac{\partial}{\partial \theta_{b}}\}_{+} =0$, offering
together $2\cdot2^d$ operators, the half of which are superposition of products of  
$\theta^{a}$ and another half corresponding superposition of 
$\frac{\partial}{\partial \theta_{a}}$.
\begin{eqnarray}
\label{thetaderanti}
\{\theta^{a}, \theta^{b}\}_{+}=0\,, \, && \,
\{\frac{\partial}{\partial \theta_{a}}, \frac{\partial}{\partial \theta_{b}}\}_{+} =0\,,
\nonumber\\
\{\theta_{a},\frac{\partial}{\partial \theta_{b}}\}_{+} &=&\delta_{ab}\,, (a,b)=(0,1,2,3,5,\cdots,d)\,.
\end{eqnarray}
Defining~\cite{nh2018} 
\begin{eqnarray}
(\theta^{a})^{\dagger} &=& \eta^{a a} \frac{\partial}{\partial \theta_{a}}\,,\nonumber\\
{\rm it \, follows}\nonumber\\
(\frac{\partial}{\partial \theta_{a}})^{\dagger}&=& \eta^{a a} \theta^{a}\,. 
\label{thetaderher}
\end{eqnarray}
The identity is the self adjoint member.
The signature $\eta^{a b}=diag\{1,-1,-1,\cdots,-1\}$ is assumed.

It appears useful to arrange $2^d$ products of $\theta^{a}$ into  irreducible
representations with 
respect to the Lorentz group with the generators~\cite{norma93} 
\begin{eqnarray}
{\cal {\bf S}}^{a b} &=& i \, (\theta^{a} \frac{\partial}{\partial \theta_{b}} - \theta^{b}
\frac{\partial}{\partial \theta_{a}})\,, \quad  ({\cal {\bf S}}^{a b})^{\dagger} = \eta^{a a}
\eta^{b b} {\cal {\bf S}}^{a b}\, . 
\label{thetasab}
\end{eqnarray} 
%
$2^{d-1}$ members of the representations  have an odd Grassmann character (those which are 
superposition of  odd products of $\theta^{a}$'s).  All the members of any particular odd 
irreducible representation follow from any starting member by the application of 
${\cal {\bf S}}^{a b}$'s.

If we exclude the self adjoint identity there is $(2^{d-1}-1)$ members of an even Grassmann 
character, they are even products of $\theta^{a}$'s. All the members of any particular even 
representation follow from any starting member by the application of ${\cal {\bf S}}^{a b}$'s.

The Hermitian conjugated $2^{d-1}$ odd partners of  odd representations of $\theta^a$'s 
and $(2^{d-1}-1)$  even partners of even representations of $\theta^a$'s are reachable from 
odd and even representations, respectively, by the application of Eq.~(\ref{thetaderher}).

It  appears useful as well to make the choice of the Cartan subalgebra of the commuting operators of the Lorentz algebra as follows
\begin{eqnarray}
{\cal {\bf S}}^{03}, {\cal {\bf S}}^{12}, {\cal {\bf S}}^{56}, \cdots, 
{\cal {\bf S}}^{d-1 \;d}\,, 
\label{cartangrass}
\end{eqnarray}
and choose the members of the irreducible representations of the Lorentz group to be the 
eigenvectors of all  the members of the Cartan subalgebra of Eq.~(\ref{cartangrass})  
\begin{eqnarray}
{\cal {\bf S}}^{ab} \,\frac{1}{\sqrt{2}}\, (\theta^a + \frac{\eta^{aa}}{i k} \theta^b) &=&
k\,\frac{1}{\sqrt{2}} (\theta^a + \frac{\eta^{aa}}{ik} \theta^b) \,, \nonumber\\
{\cal {\bf S}}^{ab} \,\frac{1}{\sqrt{2}}\, (1+ \frac{i}{k}  \theta^a \theta^b) &=&0\,,
 \nonumber\\ &{\rm or} &\;  \nonumber\\
{\cal {\bf S}}^{ab} \,\frac{1}{\sqrt{2}}\,  \frac{i}{k}  \theta^a \theta^b &=&0\,,
\label{eigengrasscartan}
\end{eqnarray}
with $k^2=\eta^{aa}\eta^{bb}$.  The  eigenvector $\frac{1}{\sqrt{2}}\, 
(\theta^0 \mp  \theta^3)$ of   ${\cal {\bf S}}^{03}$ has the eigenvalue $k=\pm i$,
the eigenvalues of all the other eigenvectors of the rest of the Cartan subalgebra 
members, Eq.~(\ref{cartangrass}), are  $k=\pm 1$.

We choose the "basis vectors" to be products of  odd nilpotents 
$\frac{1}{\sqrt{2}}\, (\theta^a + \frac{\eta^{aa}}{i k} \theta^b)$ and  of even objects $  \frac{i}{k}  \theta^a \theta^b $, with eigenvalues $k=\pm i$
 and $0$, respectively.  
 
Let us check how does ${\cal {\bf S}}^{ac}= i (\theta^a \frac{\partial}{\partial \theta_c} - 
\theta^c \frac{\partial}{\partial \theta_a})$ transform the product of two "nilpotents"
$\frac{1}{\sqrt{2}}\, (\theta^a + \frac{\eta^{aa}}{i k} \theta^b)$ and 
$\frac{1}{\sqrt{2}}\, (\theta^c + \frac{\eta^{cc}}{i k'} \theta^d)$.
Taking into account Eq.~(\ref{thetasab}) one finds that ${\cal S}^{ac}$ 
$\frac{1}{\sqrt{2}}\, (\theta^a + \frac{\eta^{aa}}{i k} \theta^b)$  
$\frac{1}{\sqrt{2}}\, (\theta^c + \frac{\eta^{cc}}{i k'} \theta^d)$ $ =
 - \frac{\eta^{aa} \eta^{cc}}{2k}$ $(\theta^a \theta^b  + \frac{k}{k'} \theta^c \theta^d)$.
 ${\cal S}^{ac}$ transforms the product of   two Grassmann odd  eigenvectors of the Cartan 
subalgebra into the superposition of two Grassmann even eigenvectors. 
 

"Basis vectors" have an odd or an even Grassmann character, if their products contain 
an odd  or an even number of "nilpotents", $\frac{1}{\sqrt{2}}\, (\theta^a + 
\frac{\eta^{aa}}{i k} \theta^b)$, respectively. 
"Basis vectors" are normalized, up to a phase, in accordance with Eq.~(\ref{grassintegral}) of  	\ref{normgrass}.


The Hermitian conjugated representations of (either an odd or an even) products of $\theta^a$'s 
can be obtained by taking into account  Eq.~(\ref{thetaderher}) for each "nilpotent"  
\begin{eqnarray}
\frac{1}{\sqrt{2}} (\theta^a + \frac{\eta^{aa}}{i k} \theta^b)^{\dagger} &=&
\eta^{aa}\,\frac{1}{\sqrt{2}} (\frac{\partial}{\partial \theta_{a}} + \frac{\eta^{aa}}{- i k}
 \frac{\partial}{\partial \theta_{b}}) \,, \nonumber\\
  (\frac{i}{k}\, \theta^a \theta^b)^{\dagger} &=& 
\frac{i}{k} \,\frac{\partial}{\partial \theta_{a}}\,\frac{\partial}{\partial \theta_{b}}\,.
\label{grasscartanher}
\end{eqnarray}
%

Making a choice of the identity for the vacuum state, 
\begin{eqnarray}
\label{vactheta}
|\phi_{og}> &=& |\,1>\,, 
\end{eqnarray}
we see that algebraic products --- we shall use  a dot ,${\,}\cdot{\,}$, or without  a dot 
for an algebraic product of eigenstates of the Cartan subalgebra forming ''basis vectors''
and $*_{A}$ for the algebraic product of  ''basis vectors''  --- of different $\theta^a$'s, 
if applied on such a vacuum state, give always nonzero contributions, \\
$$ (\theta^0 \mp \theta^3) \cdot (\theta^1 \pm i \theta^2)
\cdots  (\theta^{d-1} \mp \theta^d) |\,1>\ne {\rm zero},$$ 
(this is true also, if we substitute any of  nilpotents 
$ \frac{1}{\sqrt{2}} (\theta^a + \frac{\eta^{aa}}{i k} \theta^b)$ or all of them 
with the corresponding  even operators $(\frac{i}{k}\, \theta^a \theta^b)$; in the case
of odd Grassmann irreducible representations at least one nilpotent must remain).
The Hermitian conjugated partners, Eq.~(\ref{grasscartanher}), applied on $|\,1>$, give 
always zero  
$$ (\frac{\partial}{\partial \theta_{0}} \mp 
\frac{\partial}{\partial \theta_{3}}) \cdot (\frac{\partial}{\partial \theta_{1}} \pm i 
\frac{\partial}{\partial \theta_{2}})\cdots ( \frac{\partial}{\partial \theta_{d-1}} \pm i 
\frac{\partial}{\partial \theta_{d}}) |\,1> = 0.$$ 

Let us notice the properties of the odd products $\theta^a$'s and of their Hermitian 
conjugated partners:\\
 {\bf i.}  Superposition of  
 products of different  $\theta^a$'s,  applied on the vacuum state  
$|\,1>$, give nonzero contribution.  To create on the vacuum state the  ''fermion'' states
we make a choice of the ''basis vectors'' of the odd number of $\theta^a$'s, arranging them
to be the eigenvectors of all the Cartan subalgebra elements, Eq.~(\ref{cartangrass}).   \\
{\bf ii.}  The Hermitian conjugated partners of the ``basis vectors'', they are 
products of derivatives $\frac{\partial}{\partial \theta_{a}}$'s,  give, when applied  
on the vacuum state  $|\,1>$, Eq.~(\ref{vactheta}),  zero. Each annihilation operator 
annihilates the corresponding creation operator.\\
{\bf iii.} The algebraic product, $*_{A}$, of  a ``basis vector'' by itself gives zero, 
the algebraic anticommutator of any two ''basis vectors'' of an odd Grassmann character
(superposition of an odd products of $\theta^a$'s) gives zero 
(''basis vectors'' of the two decuplets in Table~\ref{Table grassdecuplet.} and the ''basis vector'' of Eq.~(\ref{nondecuplet}) $\frac{1}{2} (\theta^0\mp\theta^3)$, for example, demonstrate
this property).\\
 {\bf iv.}  The algebraic  application of  any annihilation operator on the corresponding Hermitian conjugated ''basis vector'' gives identity, on all the rest of  ''basis vectors'' gives
  zero. Correspondingly the algebraic  anticommutators of the creation operators and their Hermitian conjugated partners, applied on the vacuum state, give identity,  all the rest anticommutators of creation and annihilation operators applied on the vacuum state, 
  give zero.\\
{\bf v.} Correspondingly the ``basis vectors'' and their Hermitian conjugated 
partners, applied on the vacuum state $|\,1>$, Eq.~(\ref{vactheta}), fulfill the properties of  creation  and annihilation operator, respectively, for the second quantized ''fermions''
on the level of one ''fermion'' state. \\

\subsection{ Grassmann "basis vectors"}
\label{propertiesGrass1}

We construct  $ 2^{d-1}$ Grassmann odd "basis vectors" and $ 2^{d-1}-1$ (we skip 
self adjoint identity, which we use to describe the vacuum state $|\,1>$)  Grassmann 
even "basis vectors" as superposition of odd and even products of $\theta^a$'s,
respectively. Their Hermitian conjugated $ 2^{d-1}$ odd and 
$ 2^{d-1}-1$ even partners are, according to Eqs.~(\ref{thetaderher}, \ref{grasscartanher}), determined by the corresponding superposition of  odd and even  products of  $\frac{\partial}{\partial \theta_{a}}$'s, respectively~\footnote{  
Relations among operators and their Hermitian conjugated partners in both kinds of 
the Clifford algebra objects are more complicated than in the Grassmann case, where the 
Hermitian conjugated operators follow by taking into account Eq.~(\ref{thetaderher}).
In the Clifford case 
$\frac{1}{2} (\gamma^a + \frac{\eta^{aa}}{i\,k} \gamma^b)^{\dagger}$  is 
proportional to $\frac{1}{2} (\gamma^a + \frac{\eta^{aa}}{i \,(- k)} \gamma^b) $, 
while $ \frac{1}{\sqrt{2}} (1 +  \frac{i}{k}  \gamma^a \gamma^b)$ are self adjoint. 
This is the case also for representations in the sector of $\tilde{\gamma}^a$'s.}. 

\vspace{5mm}

$\;\;{\it A.a.}$ {\it Grassmann anticommuting "basis vectors" with integer spins}

\vspace{2mm}

%

Let us choose  in $d=2(2n+1)$-dimensional space-time, $n$ is a positive integer, the starting 
Grassmann odd "basis vector" $\hat{b}^{\theta 1\dagger}_{1}$, which is the
eigenvector of the Cartan subalgebra of Eqs.~(\ref{cartangrass}, \ref{eigengrasscartan}) 
with the egenvalues
 $(+i, +1, +1, \cdots, +1 )$, respectively, and has the Hermitian conjugated partner 
 equal to $(\hat{b}^{\theta 1\dagger}_{1})^{\dagger}=$ $\hat{b}^{\theta 1}_{1}$,
\begin{eqnarray}
&&\hat{b}^{\theta 1\dagger}_{1}{\bf :}= (\frac{1}{\sqrt{2}})^{\frac{d}{2}} \,
(\theta^0 - \theta^3) (\theta^1 + i \theta^2) (\theta^5 + i\theta^6) \nonumber\\
&& {} \cdots (\theta^{d-1} + i \theta^{d}) \,, \nonumber\\
&& \hat{b}^{\theta 1}_{1}  {\bf :} =
 (\frac{1}{\sqrt{2}})^{\frac{d}{2}}\,
 (\frac{\partial}{\partial \theta^{d-1}} - i \frac{\partial}{\partial \theta^{d}})
{}\cdots (\frac{\partial}{\partial \theta^{0}}
-\frac{\partial}{\partial \theta^3})\,.
\label{start(2n+1)2theta}
\end{eqnarray}
%

In the case of $d=4n$, $n$ is a positive integer, the corresponding starting Grassmann odd "basis vector" 
can be chosen as  
\begin{eqnarray}
{\hat b}^{\theta 1 \dagger}_{1}{\bf :} &=&(\frac{1}{\sqrt{2}})^{\frac{d}{2}-1} \,
(\theta^0 - \theta^3) (\theta^1 + i \theta^2) (\theta^5 + i \theta^6)\cdots
                              \nonumber\\
  &&{}\cdots (\theta^{d-3} +
 i \theta^{d-2})  \theta^{d-1} \theta^d\,.
\label{start4ntheta}
\end{eqnarray}
All the rest of "basis vectors", belonging to the same irreducible representation of the 
Lorentz group, follow by the application of ${\cal {\bf S}}^{ab}$'s.
 
We denote the members $i$ of this starting irreducible representation $k$ by 
$\hat{b}^{\theta k \dagger}_{i}$ 
and their Hermitian conjugated partners by $\hat{b}^{\theta k }_{i}$, with $k=1$.

 "Basis vectors", belonging to different irreducible representations $k=2$, will be 
denoted by $\hat{b}^{\theta 2 \dagger}_{j}$ and their Hermitian conjugated 
partners by 
$\hat{b}^{\theta 2 }_{j}=(\hat{b}^{\theta k \dagger}_{j})^{\dagger} $.

${\cal {\bf S}}^{ac}$'s, which do not belong to the Cartan subalgebra, transform 
step by step the two by two "nilpotents", no matter how many "nilpotents" are 
between the chosen two, up to a constant,  
as follows:\\ 
${\cal {\bf S}}^{ac}
\frac{1}{\sqrt{2}}\, (\theta^a + \frac{\eta^{aa}}{i k} \theta^b) \cdots $ 
$\frac{1}{\sqrt{2}}\, (\theta^c + \frac{\eta^{cc}}{i k'} \theta^d)$
$\propto - \frac{\eta^{aa} \eta^{cc}}{2k} (\theta^a \theta^b  + 
\frac{k}{k'} \theta^c \theta^d) \cdots$,\\
leaving at each step at least one "nilpotent" unchanged, so that the whole irreducible
representation remains odd.\\ 
The superposition of ${\cal {\bf S}}^{bd}$ and $ i {\cal {\bf S}}^{bc}$
transforms $- \frac{\eta^{aa} \eta^{cc}}{2k}$ $(\theta^a \theta^b  +
 \frac{k}{k'} \theta^c \theta^d)$ into $\frac{1}{\sqrt{2}} $
$(\theta^a - \frac{\eta^{aa}}{i k} \theta^b) 
 \frac{1}{\sqrt{2}}\, (\theta^c - \frac{\eta^{cc}}{i k'} \theta^d)$, and not into
$\frac{1}{\sqrt{2}}\, (\theta^a + \frac{\eta^{aa}}{i k} \theta^b)$ 
$\frac{1}{\sqrt{2}}\, (\theta^c - \frac{\eta^{cc}}{i k'} \theta^d)$ or into
$\frac{1}{\sqrt{2}}\, (\theta^a - \frac{\eta^{aa}}{i k} \theta^b)$ 
$\frac{1}{\sqrt{2}}\, (\theta^c + \frac{\eta^{cc}}{i k'} \theta^d)$.

Therefore we can start another odd representation with the "basis vector" 
$\hat{b}^{\theta 2 \dagger }_{1}$  as follows
%
\begin{eqnarray}
\hat{b}^{\theta 2\dagger}_{1}{\bf :} &=& (\frac{1}{\sqrt{2}})^{\frac{d}{2}} \,
(\theta^0 + \theta^3) (\theta^1 + i \theta^2) (\theta^5 + i\theta^6) 
 {} \cdots (\theta^{d-1} + i \theta^{d}) \,, \nonumber\\
(\hat{b}^{\theta 2 \dagger}_{1})^{\dagger} &=& \hat{b}^{\theta 1}_{2} {\bf :} =
 (\frac{1}{\sqrt{2}})^{\frac{d}{2}}\,
 (\frac{\partial}{\partial \theta^{d-1}} - i \frac{\partial}{\partial \theta^{d}})
{}\cdots (\frac{\partial}{\partial \theta^{0}}
- \frac{\partial}{\partial \theta^3})\,.
\label{start(2n+1)2thetasecond}
\end{eqnarray}
The application of ${\cal {\bf S}}^{ac}$'s determines the whole second irreducible representation 
$\hat{b}^{\theta 2 \dagger }_{j}$.
 
One finds that each of these two irreducible representations has 
${\bf \frac{1}{2}\frac{d!}{\frac{d}{2}! \frac{d}{2}!}}$ members, Ref.~\cite{nh2018}.



%
%
Taking into account Eq.~(\ref{thetaderanti}), it follows that odd products of $\theta^a$'s 
anticommute and so do the odd products of $\frac{\partial}{\partial \theta_{a}}$'s.

{\bf Statement 1: } The oddness of the products of  $\theta^a$'s guarantees the 
anticommuting properties of all objects which include odd number of $\theta^a$'s.

One further sees that $\frac{\partial}{\partial \theta^{a}} \theta^b = \eta^{ab}$,
while $\frac{\partial}{\partial \theta_{a}}|\,1>=0$, and $\theta^a |\,1>=\theta^a  |\,1>$.
and $\{ \hat{b}^{\theta k}_i, \hat{b}^{\theta l \dagger}_{j} \}_{*_{A}+}= $
We can therefore conclude
\begin{eqnarray}
\{ \hat{b}^{\theta k}_i, \hat{b}^{\theta l \dagger}_{j} \}_{*_{A}+} 
|\,\,1> &=& \delta_{i j}\; \delta^{k l}\;|\,\,1>\,,\nonumber\\
\{ \hat{b}^{\theta k}_i, \hat{b}^{\theta l}_{j} \}_{*_{A}+}  |\,\,1>
&=& 0\;\cdot\, |\,\,1> \,,\nonumber\\
\{\hat{b}^{\theta k \dagger}_i,\hat{b}^{\theta l \dagger}_{j}\}_{*_{A}+} \;|\,\,1>
&=&0\;\cdot\, |\,\,1> \,,\nonumber\\
\hat{b}^{\theta k}_{j} \,*_{A}\,|\,\,1>& =&0\;\cdot\, |\,\,1> \,, 
\label{ijthetaprod}
\end{eqnarray}
where $\{ \hat{b}^{\theta k}_i, \hat{b}^{\theta l \dagger}_{j} \}_{*_{A}+}=$
$ \hat{b}^{\theta k}_i *_{A} \hat{b}^{\theta l \dagger}_{j} + \hat{b}^{\theta l}_j  *_{A}\hat{b}^{\theta k \dagger}_{i} $ is meant.
 
These anticommutation relations of the ''basis vectors'' of the odd Grassmann character, 
manifest on the level of the Grassmann algebra the  anticommutation relations 
required by Dirac~\cite{Dirac} for second quantized fermions.

The ''Grassmann fermion basis states'' can be obtained by the application of 
creation operators $\hat{b}^{\theta k \dagger}_{i}$
on the vacuum state $ |\,1>$
\begin{eqnarray}
 |\phi^{k}_{o\, i }> &=& \hat{b}^{\theta k\dagger}_{i} \,|\,1>.
\label{ijthetaphi}
\end{eqnarray}
  We use them to determine the internal space of ''Grassmann fermions'' in the tensor 
  product $*_{T}$ of these ''basis states'' and of the momentum space, when 
  looking for the anticommuting  single particle ''Grassmann states'', which have,  according to Eq.~(\ref{eigengrasscartan}), an integer spin,
and not half integer spin as it is the case for  the so far observed fermions.

%

\vspace{5mm}

$\;\;{\it A.b.}$ {\it Illustration of anticommuting ''basis vectors'' in $d=(5+1)$-dimensional space}

\vspace{2mm}

\begin{small}
Let us illustrate  properties of Grassmann odd representations for  
$d=(5+1)$-dimensional space. 

Table~\ref{Table grassdecuplet.} represents two decuplets, which are 
"egenvectors" of the Cartan subalgbra (${\cal {\bf S}}^{03}$,  ${\cal {\bf S}}^{12}$, 
${\cal {\bf S}}^{56}$), Eq.~(\ref{cartangrass}), of the Lorentz algebra ${\cal {\bf S}}^{ab}$. 
The two decuplets represent two Grassmann odd  irreducible representations of $SO(5,1)$. 

One can read on the same table, from the first to the third and from the fourth to the sixth line 
of both decuplets, two Grassmann even triplet representations of $SO(3,1)$, if paying attention 
on the eigenvectors
of ${\cal {\bf S}}^{03}$ and  ${\cal {\bf S}}^{12}$ alone,  while the  eigenvector of 
${\cal {\bf S}}^{56}$ has, as a "spectator", the eigenvalue  either $+1$ (the first triplet in both 
decuplets) or $-1$ (the second triplet in both decuplets).  Each of the two decuplets contains 
also one "fourplet"  with the "charge" ${\cal {\bf S}}^{56}$ equal to zero ($(7^{th}, 8^{th}, 9^{th}, 
10^{th})$ lines in each of the two decuplets (Table II in Ref.~\cite{norma93})).  

Paying attention on the eigenvectors of ${\cal {\bf S}}^{03}$ alone one recognizes as 
well even and odd representations of $SO(1,1)$: $\theta^0 \theta^3$ 
and $\theta^0 \pm \theta^3$, respectively.

The Hermitian conjugated "basis vectors" follow  by using $\,$ Eq.~(\ref{grasscartanher}) and 
is for the first "basis vector" of Table~\ref{Table grassdecuplet.} equal to $(-)^2 
(\frac{1}{\sqrt{2}})^3 (\frac{\partial}{\partial \theta_5} -i
\frac{\partial}{\partial \theta_6})\, (\frac{\partial}{\partial \theta_1} - i
\frac{\partial}{\partial \theta_2})\, (\frac{\partial}{\partial \theta_0} + 
\frac{\partial}{\partial \theta_3})$. One correspondingly finds that when $(\frac{1}{\sqrt{2}})^3 
(\frac{\partial}{\partial \theta_5} -i \frac{\partial}{\partial \theta_6})\, (\frac{\partial}{\partial \theta_1} - i
\frac{\partial}{\partial \theta_2})\, (\frac{\partial}{\partial \theta_0} + 
\frac{\partial}{\partial \theta_3}$) applies on $(\frac{1}{\sqrt{2}})^3
 (\theta^{0} - \theta^{3}) (\theta^{1} + i \theta^{2}) (\theta^{5} + i \theta^{6}) $ the result is 
identity. Application of $ (\frac{1}{\sqrt{2}})^3 (\frac{\partial}{\partial \theta_5} -i
\frac{\partial}{\partial \theta_6})\, (\frac{\partial}{\partial \theta_1} - i
\frac{\partial}{\partial \theta_2})\, (\frac{\partial}{\partial \theta_0} + 
\frac{\partial}{\partial \theta_3})$ on all the rest of "basis vectors" of the decuplet $I$ as well as 
on all the "basis vectors" of the decuplet $II$ gives zero. "Basis vectors" are orthonormalized with respect to
 Eq.~(\ref{grassintegral}).
%
 \begin{table*}
 \caption{\label{Table grassdecuplet.} The two decuplets, the odd eigenvectors of the 
Cartan subalgebra, Eq.~(\ref{cartangrass}), (${\cal {\bf S}}^{0 3}, {\cal {\bf S}}^{1 2}$, 
${\cal {\bf S}}^{5 6}$, for $SO(5,1)$) of the Lorentz algebra in Grassmann 
$(5+1)$-dimensional space, forming two irreducible representations, are presented.
Table is partly taken from Ref.~\cite{nh2018}.  
 The "basis vectors" within each decuplet are reachable from any member by 
${\cal {\bf S}}^{ab}$'s and are decoupled from another decuplet.
The two operators of handedness,  $\Gamma^{((d-1)+1)}$ for $d=(6,4)$, are invariants of the 
Lorentz algebra, Eq.~(\ref{handedness}), $\Gamma^{(5+1)}$ for the whole decuplet, 
$\Gamma^{(3+1)}$ for the "triplets" and "fourplets".}
\begin{tiny}
 \begin{tabular}{|c|r|r|r|r|r|r|r|}
 \hline
$I$&$i$ &$\rm{decuplet\; of\; eigenvectors} $&${\cal {\bf S}}^{03}$&${\cal {\bf S}}^{1 2}$&
${\cal {\bf S}}^{5 6}$&$\Gamma^{(5+1)}$&$\Gamma^{(3+1)}$\\
 \hline 
& $1$  & ($\frac{1}{\sqrt{2}})^3 (\theta^{0} - \theta^{3}) (\theta^{1} + i \theta^{2})
 (\theta^{5} + i \theta^{6})$ &$ i$&$ 1$&$1$&$1$&$1$\\
\hline
&$2$  & ($\frac{1}{\sqrt{2}})^2 (\theta^{0} \theta^{3} + i \theta^{1} \theta^{2}) 
 (\theta^{5} + i \theta^{6})$ & $ 0$ & $0 $ &$1$&$1$&$1$\\
\hline
&$3$  & ($\frac{1}{\sqrt{2}})^3 (\theta^{0} +  \theta^{3}) (\theta^{1} - i \theta^{2})
  (\theta^{5} + i \theta^{6})$ &$-i $&$-1$&$1$&$1$&$1$\\
\hline
&$4$  & ($\frac{1}{\sqrt{2}})^3 (\theta^{0} -  \theta^{3}) (\theta^{1} - i \theta^{2})
  (\theta^{5} - i \theta^{6})$ &$ i $&$-1$&$-1$&$1$&$-1$\\
\hline
&$5$  & ($\frac{1}{\sqrt{2}})^2 (\theta^{0} \theta^{3} - i \theta^{1} \theta^{2}) 
 (\theta^{5} - i \theta^{6})$ & $ 0 $& $0 $&$-1$&$1$&$-1$\\
\hline
&$6$  & ($\frac{1}{\sqrt{2}})^3 (\theta^{0} +  \theta^{3}) (\theta^{1} + i \theta^{2})
  (\theta^{5} - i \theta^{6})$ &$-i $&$ 1$&$-1$&$1$&$-1$\\
\hline
& $7$  & ($\frac{1}{\sqrt{2}})^2 (\theta^{0} - \theta^{3}) (\theta^{1} \theta^{2} +
 \theta^{5} \theta^{6})$ &$ i$&$ 0$&$0$&$1$&$0$\\
\hline
& $8$  & ($\frac{1}{\sqrt{2}})^2 (\theta^{0} + \theta^{3}) (\theta^{1} \theta^{2} -
 \theta^{5} \theta^{6})$ &$- i$&$ 0$&$0$&$1$&$0$\\
\hline
& $9$  & ($\frac{1}{\sqrt{2}})^2 (\theta^{0}  \theta^{3} +i \theta^{5} \theta^{6}) 
(\theta^{1}+i \theta^{2}) 
$ &$ 0$&$ 1$&$0$&$1$&$0$\\
\hline
& $10$  & ($\frac{1}{\sqrt{2}})^2 (\theta^{0}  \theta^{3} - i \theta^{5} \theta^{6}) 
(\theta^{1}-i \theta^{2}) 
$ &$ 0$&$- 1$&$0$&$1$&$0$\\
\hline\hline 
$II$&$i$ &$\rm{decuplet\; of\; eigenvectors}$&${\cal {\bf S}}^{03}$&${\cal {\bf S}}^{1 2}$&
${\cal {\bf S}}^{5 6}$&$\gamma^{(5+1)}$&$\gamma^{(3+1)}$\\ 
 \hline 
& $1$  & ($\frac{1}{\sqrt{2}})^3 (\theta^{0} + \theta^{3}) (\theta^{1} + i \theta^{2})
 (\theta^{5} + i \theta^{6})$ &$- i$&$ 1$&$1$&$-1$&$-1$\\
\hline
&$2$  & ($\frac{1}{\sqrt{2}})^2 (\theta^{0} \theta^{3} - i \theta^{1} \theta^{2}) 
 (\theta^{5} + i \theta^{6})$ & $ 0$&$ 0 $&$1$&$-1$&$-1$\\
\hline
&$3$  & ($\frac{1}{\sqrt{2}})^3 (\theta^{0} -  \theta^{3}) (\theta^{1} - i \theta^{2})
  (\theta^{5} + i \theta^{6})$ &$ i $&$-1$&$1$&$-1$&$-1$\\
\hline
&$4$  & ($\frac{1}{\sqrt{2}})^3 (\theta^{0} +  \theta^{3}) (\theta^{1} - i \theta^{2})
  (\theta^{5} - i \theta^{6})$ &$- i $&$-1$&$-1$&$-1$&$1$\\
\hline
&$5$  & ($\frac{1}{\sqrt{2}})^2 (\theta^{0} \theta^{3} + i \theta^{1} \theta^{2}) 
 (\theta^{5} - i \theta^{6})$ & $ 0$& $0 $&$-1$&$-1$&$1$\\
\hline
&$6$  &($\frac{1}{\sqrt{2}})^3 (\theta^{0} -  \theta^{3}) (\theta^{1} + i \theta^{2})
  (\theta^{5} - i \theta^{6})$ &$ i $&$ 1$&$-1$&$-1$&$1$\\
\hline
& $7$  & ($\frac{1}{\sqrt{2}})^2 (\theta^{0} + \theta^{3}) (\theta^{1} \theta^{2} +
 \theta^{5} \theta^{6})$ &$- i$&$ 0$&$0$&$-1$&$0$\\
\hline
& $8$  & ($\frac{1}{\sqrt{2}})^2 (\theta^{0} - \theta^{3}) (\theta^{1} \theta^{2} -
 \theta^{5} \theta^{6})$ &$ i$&$ 0$&$0$&$-1$&$0$\\
\hline
& $9$  & ($\frac{1}{\sqrt{2}})^2 (\theta^{0} \theta^{3} - i \theta^{5} \theta^{6}) 
(\theta^{1}+i \theta^{2}) 
$ &$ 0$&$ 1$&$0$&$-1$&$0$\\
\hline
& $10$  & ($\frac{1}{\sqrt{2}})^2 (\theta^{0}  \theta^{3} + i \theta^{5} \theta^{6}) 
(\theta^{1}-i \theta^{2}) 
$ &$ 0$&$- 1$&$0$&$-1$&$0$\\
\hline\hline 
 \end{tabular}
\end{tiny}
%
 \end{table*} 
Let us notice that $\frac{\partial}{\partial \theta_{a}}$ on a "state" which is just  an identity, $|\,1>$,
gives zero, $\frac{\partial}{\partial \theta_{a}}\,|\,1>=0$, while $\theta^a \,|\,1>$, or 
any superposition of products of $\theta^a$'s, applied on $|\,1>$, gives the "vector" back.

One easily sees that application of 
products of superposition of $\theta^a$'s on $|\,1>$  gives nonzero contribution, while 
application of products of superposition of $\frac{\partial}{\partial \theta^a}$'s on $|\,1>$ gives 
zero.

The two by ${\cal {\bf S}}^{ab}$ decoupled Grassmann decuplets of 
Table~\ref{Table grassdecuplet.} are the largest two 
irreducible representations of  odd products of $\theta^a$'s. There are $12$  additional Grassmann 
odd "vectors", arranged into irreducible representations of six singlets and six sixplets
\begin{eqnarray} 
&& (\frac{1}{2}\,(\theta^0\mp \theta^3), \frac{1}{2}\,(\theta^1\pm i \theta^2), 
 \frac{1}{2}\,(\theta^5 \pm i \theta^6)\,,\nonumber\\
&&\frac{1}{2}\,(\theta^0 \mp \theta^3)\, \theta^1 \theta^2 \theta^5 \theta^6,
\frac{1}{2}\,(\theta^1\pm i \theta^2)\, \theta^0 \theta^3 \theta^5 \theta^6\,, 
\frac{1}{2}\,(\theta^5 \pm i \theta^6)\,\theta^0 \theta^3 \theta^1 \theta^2)\,.
\label{nondecuplet}
\end{eqnarray}

The algebraic application of products of superposition of 
$\frac{\partial}{\partial \theta^a}$'s on the corresponding Hermitian conjugated partners, which are products of superposition of $\theta^a$'s,
leads to the identity for either even or odd Grassmann character~\footnote{
We shall see in Part II that the vacuum states are in the Clifford case, similarly as in the Grassmann case, for both kinds of the Clifford algebra objects, $\gamma^a$'s and
 $\tilde{\gamma}^a$'s, sums of products of the annihilation $\times$ its Hermitian conjugated creation operators, and correspondingly self adjoint operators, 
but they are not the identity.}.

Besides $32$ Grassmann odd eigenvectors of the Grassmann Cartan subalgebra, 
Eq.~(\ref{cartangrass}), there are $(32-1)$ Grassmann "basis vectors", which we  arrange into irreducible representations, which are superposition of even products of 
$\theta^a$'s. The even self adjoint operator identity 
(which is indeed the normalized product of all the annihilation times $*_{A}$ creation operators)  is used to represent the vacuum state. 

%

%


It is not difficult to see that  Grassmann "basis vectors" of an odd Grassmann character 
anticommute among themselves and so do odd products of superposition of 
$\frac{\partial}{\partial \theta^a}$'s, while equivalent even
products commute. 



The Grassmann odd algebra (as well as the two odd Clifford algebras) offers, due to the
oddness of the internal space giving oddness as well to the elements of the tensor products
of the internal space and of the momentum space, the 
description of the anticommuting second quantized fermion fields, as postulated by Dirac. 
But  the Grassmann "fermions" carry the integer spins, while the observed fermions ---  
quarks and leptons --- carry half integer spin.
\end{small}
%

%
\vspace{5mm}

$\;\;{\it A.c.}$ {\it Grassmann commuting "basis vectors" with integer spins}

\vspace{2mm}

\begin{small}

Grassmann even  "basis vectors" manifest the commutation relations, and not the anticommutation ones as it is the case for the Grassmann odd "basis vectors".
Let us use in the Grassmann even case, that is the case of superposition of an even 
number of $\theta^a$'s  
in $d=2(2n+1)$, the notation $\hat{a}^{\theta k \dagger}_j $, again chosen to be
eigenvectors of the Cartan subalgebra, Eq.~(\ref{cartangrass}),  and let us start with 
one  representative
\begin{eqnarray}
\hat{a}^{\theta 1 \dagger}_j {\bf :} &=&(\frac{1}{\sqrt{2}})^{\frac{d}{2}-1} \,
  (\theta^0 - \theta^3) (\theta^1 + i \theta^2) (\theta^5 + i \theta^6)\nonumber\\
   &&{}\cdots (\theta^{d-3} +
 i \theta^{d-2})  \theta^{d-1} \theta^d\,.
\label{start2(2n+1)thetaeven}
\end{eqnarray}
The rest of "basis vectors", belonging to the same Lorentz irreducible representation, follow by 
the application of ${\cal \bf{S}}^{ab}$. The Hermitian conjugated partner of 
$\hat{a}^{\theta 1 \dagger}_1$ is $\hat{a}^{\theta 1}_1 =
 (\hat{a}^{\theta 1 \dagger}_1)^{\dagger}$
\begin{eqnarray}
\hat{a}^{\theta 1}_{1} {\bf :} &=& (\frac{1}{\sqrt{2}})^{\frac{d}{2}-1}\,
\frac{\partial}{\;\partial \theta^{d}}
\frac{\partial}{\;\partial \theta^{d-1}} (\frac{\partial}{\;\partial \theta^{d-3}} -
                           i \frac{\partial}{\;\partial \theta^{d-2}})\nonumber\\
  &&{}\cdots (\frac{\partial}{\;\partial \theta^{0}}
-\frac{\partial}{\;\partial \theta^3})\,.
\label{start2(2n+1)thetaevenher}
\end{eqnarray}
%

If $\hat{a}^{\theta k \dagger}_{j}$ represents a Grassmann even creation operator, 
with index $k$ denoting different irreducible representations and index $j$ denoting a particular 
member of the  $k^{th}$ irreducible representation, while $\hat{a}^{\theta k }_{j}$ represents 
its Hermitian conjugated partner, one obtains by taking into account Sect.~\ref{propertiesGrass0},
the relations 
\begin{eqnarray}
\{ \hat{a}^{\theta k}_i, \hat{a}^{\theta k{'} \dagger}_{j} \}_{*_{A}-} 
|\,1> &=& \delta_{i j}\; \delta^{k k{'}}\;|\,1>\,,\nonumber\\
\{ \hat{a}^{\theta k}_i, \hat{a}^{\theta k{`}}_{j} \}_{*_{A}-}  |\,1>
&=& 0\;\cdot\, |\,1> \,,\nonumber\\
\{\hat{a}^{\theta k \dagger}_i,\hat{a}^{\theta k{'} \dagger}_{j}\}_{*_{A}-} \;|\,1>
&=&0\;\cdot\,|\,1> \,,\nonumber\\
\hat{a}^{\theta k}_{i} \, *_{A}\,|\,1>& =&0\;\cdot\,|\,1> \,,\nonumber\\
\hat{a}^{\theta k \dagger}_{i} \, *_{A}\,|\,1>& =&|\phi^{k}_{e\, i }>\,.
\label{ijthetaprodeven}
\end{eqnarray}
Equivalently to the case of Grassmann odd ''basis vectors'' also here 
$\{ \hat{a}^{\theta k}_i, \hat{a}^{\theta l \dagger}_{j} \}_{*_{A}-}=$
$ \hat{a}^{\theta k}_i *_{A} \hat{a}^{\theta l \dagger}_{j} - \hat{a}^{\theta l}_j  *_{A}\hat{a}^{\theta k \dagger}_{i} $ is meant.
\end{small}

%

\subsection{Action for free massless ''Grassmann fermions" with integer spin~\cite{nh2018}}
\label{actionGrass}

\vspace{2mm}

In  the Grassmann case  the ''basis vectors'' of an odd Grassmann character, 
chosen to be the eigenvectors of the Cartan subalgebra of the Lorentz algebra in Grassmann space, Eq.~(\ref{cartangrass}), manifest the anticommutation 
relations of Eq.~(\ref{ijthetaprod}) on the algebraic level.  
  
To compare the properties of creation and annihilation operators for ''integer spin 
fermions'', for which the internal degrees of freedom are described by the odd Grassmann algebra,   
with the creation and annihilation operators postulated by Dirac for the second quantized fermions depending on the quantum numbers of the internal space of fermions and on the
momentum space, we need to define the tensor product $*_{T}$  of the odd ''Grassmann 
basis states'', described by the superposition of odd products of $\theta^a$'s 
(with the finite degrees of freedom) and of the momentum (or coordinate) space 
(with the infinite degrees of freedom), taking as the basis the tensor product of both
spaces.

{\bf Statement 2:} For deriving the anticommutation relations for the ''Grassmann fermions'',
to be compared to anticommutation relations of the second quantized fermions, we need
to define the tensor product of the Grassmann odd ''basis vectors''  and the momentum space
\begin{equation}
\label{thetaptensor}
{\rm {\bf basis}}_{(p^a, \theta^a)} = |p^a>\, *_{T} \,|\theta^a>\,.
\end{equation}

We need even more, we need to find the Lorentz invariant action for, let say, free massless "Grassmann fermions" to define such a ''basis'', that would manifest the relation 
$|p^0|=|\vec{p}|$. We follow here the suggestion of 
one of us (N.S.M.B.) from Ref.~\cite{nh2018}.
%
\begin{eqnarray}
{\cal A}_{G}\,  &=&  \int \; d^dx \;d^d\theta\; \omega \, \{\phi^{\dagger} \, \gamma^{0}_{G}
 \,\frac{1}{2}\,
\theta^a p_{a} \phi \}+ h.c.\,,\nonumber\\
\omega &=&\prod^{d}_{k=0}(\frac{\partial}{\;\,\partial \theta_k} + \theta^{k})\,,
\label{actionWeylGrass}
\end{eqnarray}
with $ \gamma^{a}_{G}=(1-2\theta^a \frac{\partial}{\partial \theta_a})$, 
$(\gamma^{a}_{G})^{\dagger}= \gamma^{a}_{G}$, for each $a=(0,1,2,3,5,\cdots,d)$.
We use the integral over $\theta^a$ coordinates with the weight function $\omega$ from 
Eq.~(\ref{grassintegral}, \ref{grassnorm}).  
Requiring the Lorentz invariance we add after $\phi^{\dagger}$ the operator $\gamma^0_{G}$,
which takes care of the Lorentz  invariance. Namely
\begin{eqnarray}
\label{Linvariancegrass}
{\cal {\bf S}}^{ab \dagger}\, (1-2\theta^0 \frac{\partial}{\partial \theta^0}) &= & 
 (1-2\theta^0 \frac{\partial}{\partial \theta^0})\,{\cal {\bf S}}^{ab}\,,\nonumber\\
{\cal {\bf S}}^{\dagger} \, (1-2\theta^0 \frac{\partial}{\partial \theta^0})&=& 
(1-2\theta^0 \frac{\partial}{\partial \theta^0})\, {\cal {\bf S}}^{-1}\,,\nonumber\\
{\cal {\bf S}} &=& e^{-\frac{i}{2} \omega_{ab} (L^{ab} + {\cal {\bf S}}^{ab})}\,,
\end{eqnarray}
while $\theta^a,  \frac{\partial}{\partial \theta_a}$ and $p^a$ transform as Lorentz vectors.

The Lagrange density is up to the surface term equal to~\footnote{
Taking into account the relations  $\gamma^a= (\theta^{a} + 
\frac{\partial}{\partial \theta_a})$, $\tilde{\gamma}^a=i \,(\theta^{a} - 
\frac{\partial}{\partial \theta_a}$), $ \gamma^{0}_{G}
= - i \eta^{aa} \gamma^a \tilde{\gamma}^{a}$ the Lagrange density can be rewritten as
${\cal L}_{G}\,  = -i \frac{1}{2} \phi^{\dagger} \, \gamma^0_{G}  \,
\tilde{ \gamma}^a\,( \hat{p}_a \phi)$  
$=  -i \frac{1}{4} \{ \phi^{\dagger} \, \gamma^0_{G}  \,
\tilde{ \gamma}^a\, \hat{p}_a \phi\, - \hat{p}_a  \phi^{\dagger}\,  \, \gamma^0_{G}  \,
\tilde{ \gamma}^a\, \phi\,\}$.}
\begin{eqnarray}
{\cal L}_{G}\,  &=&  \frac{1}{2} \phi^{\dagger} \, \gamma^0_{G} 
(\theta^a - \frac{\partial}{\partial \theta_a})\,( \hat{p}_a \phi) \nonumber\\
&=&  \frac{1}{4} \{ \phi^{\dagger} \, \gamma^0_{G}  \,
(\theta^a - \frac{\partial}{\partial \theta_a})\, \hat{p}_a \phi - \nonumber\\
&&(\hat{p}_a  \phi^{\dagger})  \gamma^0_{G} 
(\theta^a - \frac{\partial}{\partial \theta_a}) \phi\}\,,
\label{LDWeylGrass10}
\end{eqnarray}
leading to the equations of motion~\footnote{
Varying the action with respect to $\phi^{\dagger}$ and $\phi$ it follows:   
$\frac{\partial {\cal L}_{G}}{\partial \phi^{\dagger}} -  
 \hat{p}_{a} \,\frac{\partial {\cal L}_{G}}{\partial \hat{p}_a \phi^{\dagger}}  = 
0 =\frac{-i}{2} \gamma^0_{G}  \, \tilde{\gamma}^a\,\hat{p}_a\,\phi$, and  
$\frac{\partial {\cal L}_{G}}{\partial \phi} -  
 \hat{p}_{a} \,\frac{\partial {\cal L}_{G}}{\partial (\hat{p}_a \phi)}  = 0=
 \frac{i}{2}\hat{p}_a \,\phi^{\dagger} \gamma^0_{G}  \, \tilde{\gamma}^a$.}
\begin{eqnarray}
\label{Weylgrass}
\frac{1}{2}\,
\gamma^0_{G} \,(\theta^a - 
\frac{\partial}{\partial \theta_a})
\, {\hat p}_{a} \,|\phi>\,&= & 0\,,
\end{eqnarray}
as well as the the ''Klein-Gordon''  equation,
 \[(\theta^a - \frac{\partial}{\partial \theta_a}) \,{\hat p}_{a} 
\,(\theta^b - \frac{\partial}{\partial \theta_b}) \, {\hat p}_{b} \,|\phi>=0 = 
{\hat p}_a {\hat p}^a \,|\phi>.\]

The eigenstates $\phi$ of equations of motion for free massless "Grassmann fermions", Eq.~(\ref{Weylgrass}), can be found as the tensor product, Eq.(\ref{thetaptensor}) of the superposition 
 of $2^{d-1}$ Grassmann odd "basis vectors" $ {\hat b}^{\theta k \dagger}_{i}$ and 
 the momentum space, represented by plane waves, applied on the
 vacuum  state $|\,1> $. Let us remind that the ''basis vectors''  are the ''eigenstates'' 
 of the Cartan subalgebra, Eq.~(\ref{cartangrass}), fulfilling  (on the algebraic level)
 the anticommutation relations of Eq.~(\ref{ijthetaprod}).  And since the oddness 
 of the Grassmann odd ''basis vectors'' guarantees the oddness of the tensor products
 of the  internal part of ''Grassmann fermions'' and of the plane waves, we expect the equivalent 
 anticommutation relations also for the eigenstates of the Eq.~(\ref{Weylgrass}), 
 which define the single particle anticommuting states of ''Grassmann fermions''. 

The coefficients, determining the superposition, depend on momentum $p^a$, 
$a=(0,1,2,3,5,$ $\dots,d)$, $|p^0| = |\vec{p}|$,  
of the plane wave solution $e^{-i p_a x^a}$. 

Let us therefore define the new creation operators and the corresponding single 
particle ''Grassmann fermion'' states as the tensor product of two spaces, the
Grassmann odd ''basis vectors'' and the momentum space basis 
\begin{eqnarray}
{\hat{\bf b}}^{\theta k \,s \dagger} (\vec{p})& \stackrel{\mathrm{def}}{=}& 
\sum_{i} c^{k s}{}_{ i}  (\vec{p})\, \hat{b}^{\theta k \dagger}_{i} \,,
 \;\;\; \qquad |p^0| = |\vec{p}|\,,\nonumber\\
\underline{\hat{\bf b}}^{\theta k \,s \dagger}_{tot} (\vec{p}) 
& \stackrel{\mathrm{def}}{=}&
{\hat{\bf b}}^{\theta k \,s \dagger} (\vec{p}) \cdot e^{-i p_a x^a} \,, \;\; \;
|p^0| = |\vec{p}|\,,\nonumber\\
<x|\phi^{k s}_{tot} (\vec{p})> 
&=& \underline{\hat{\bf b}}^{\theta ks \dagger}_{tot}(\vec{p})\,
|\, 1>\,, \;\; \qquad |p^0| = |\vec{p}|\,,
\label{ptheta}
\end{eqnarray}
with $s$ representing different solutions of the equations of motion and $k$ different 
irreducible representations of the Lorentz group, $\vec{p}$ denotes the chosen 
vector ($p^0,\vec{p}$) in momentum space. 


One has further  
\begin{eqnarray}
\label{phiksx}
|\phi^{k s} (x^0, \vec{x})> &=& \int_{- \infty}^{+ \infty} \,\frac{d^{d-1}p}{(\sqrt{2 \pi})^{d-1}} \, \underline{\hat{\bf b}}^{\theta k s \dagger} (\vec{p})|_{|p^0| = |\vec{p}|} 
|\, 1>\, 
\end{eqnarray}

The orthogonalized states $|\phi^{k s} (\vec{p})>$ fulfill  the relation 
\begin{eqnarray}
<\phi^{k s} (\vec{p})|\phi^{k' s'} (\vec{p}{}')>&=& \delta^{k k'}\, \delta_{s s'}\,
\delta_{p p'}\,,
\;\;\quad  |p^0| = |\vec{p}|\,,\nonumber\\
<\phi^{k' s'} (x^0, \vec{x}{}')|\phi^{k s} (x^0,\vec{x})>&=& \delta^{k k'}\, \delta_{s s'}\,
\delta_{\vec{x}{}', \vec{x}}\,,
\label{ortpp'}
\end{eqnarray}
where we assumed the discretization of momenta $\vec{p}$ and coordinates $\vec{x}$.

%
%
In even dimensional spaces ($d=2(2n+1)$ and $4n$) there are $2^{d-1}$
Grassmann odd superposition of ''basis vectors'', which belong to different
irreducible representations, among them twice 
${\bf \frac{1}{2} \frac{d!}{\frac{d}{2}! \frac{d}{2}!}}$ of the kind presented in
Eqs.~(\ref{start(2n+1)2theta}, \ref{start4ntheta}) and discussed in the chapter
{\it A.b.} of the subsect.~\ref{propertiesGrass1} and in 
Table~\ref{Table grassdecuplet.} for a particular case $d=(5+1)$. The 
illustration for the superposition ${\hat {\bf b}}^{\theta k\,s \dagger} (\vec{p})$
and $ \underline{\hat {\bf b}}^{\theta k\,s \dagger}_{tot} (\vec{p}) $
is presented, again for $d=(5+1)$, in chapter {\it B.a.}.

We introduced in Eq.~(\ref{ptheta}) the creation operators 
$ \underline{\hat {\bf b}}^{\theta k\,s \dagger}_{tot} (\vec{p}) $ as the tensor 
product of the ''basis vectors'' of  Grassmann  algebra elements and the momentum 
basis. The Grassmann algebra elements transfer their oddness to the 
tensor products of these two basis. Correspondingly must 
$ \underline{\hat {\bf b}}^{\theta k\,s \dagger}_{tot} (\vec{p}) $
together with their Hermitian conjugated annihilation operators $( \underline{\hat {\bf b}}^{\theta k\,s \dagger}_{tot} (\vec{p}) )^{\dagger}=$ 
$ \underline{\hat {\bf b}}^{\theta k\,s }_{tot} (\vec{p}) $ fulfill the
the  anticommutation relations equivalent to the anticommutation relations of 
Eq.~(\ref{ijthetaprod})  
\begin{eqnarray}
\{\underline{\hat {\bf b}}^{\theta k\,s}_{tot} (\vec{p}), 
\underline{\hat {\bf b}}^{\theta k'\,s' \dagger}_{tot}\, (\vec{p}{\,}') \}_{*_{T}+} 
|\,1> &=& \delta^{k k'}\; \delta_{s s'} \delta (\vec{p} -\vec{p}{\,}')\;|\,1>\,,\nonumber\\
\{ \underline{\hat {\bf b}}^{\theta k\,s}_{tot} (\vec{p}), 
\underline{\hat {\bf b}}^{\theta k'\,s'}_{tot} (\vec{p}{\,}') \}_{*_{T}+}   |\,1>
&=& 0\;\cdot\, |\,1> \,,\nonumber\\
\{\underline{\hat {\bf b}}^{\theta k\,s \dagger}_{tot}  (\vec{p}) ,
\underline{\hat {\bf b}}^{\theta k'\,s' \dagger}_{tot} (\vec{p}{\,}') \}_{*_{T}+}  
|\,1>&=&0\;\cdot\,|\,1> \,,\nonumber\\
\underline{\hat {\bf b}}^{\theta k\,s }_{tot} (\vec{p})\,*_{T}\, |\,1>& =&0\;\cdot\,|\,1> \,, \nonumber\\
|p^0|&=& |\vec{p}|\,.
\label{ijthetaprodgen}
\end{eqnarray}
$k$ labels different irreducible representations of Grassmann odd ``basis vectors'',
$s$ labels different --- orthogonal and normalized --- solutions of equations of motion 
and $\vec{p}$ represent different momenta fulfilling the relation $(p^0)^2 = (\vec{p})^2$.
Here we allow continuous momenta and take into account that 
\begin{eqnarray}
<\,1|\underline{\hat {\bf b}}^{\theta k\,s }_{tot}(\vec{p}) *_{T}
 \underline{\hat {\bf b}}^{\theta k'\,s' \dagger}_{tot} (\vec{p}{\,}') |\,1 >
 &=& \delta^{k k'} \delta^{s s'} \delta(\vec{p}- \vec{p}{\,}') \,,
\label{ortpp'con}
\end{eqnarray}
in the case of continuous values of $\vec{p}$ in even $d$-dimensional space.

For each momentum $\vec{p}$ there are $2^{d-1}$ members of the odd 
Grassmann character, belonging to different irreducible representations.
The plane wave solutions, belonging to different $\vec{p}$, are orthogonal,
defining correspondingly $\infty$ many degrees of freedom for each of 
$2^{d-1}$ ''fermion'' states, defined by 
$ \underline{\hat {\bf b}}^{\theta k\,s\dagger }_{tot} (\vec{p})$, when applying  on the vacuum state $|\,1>, Eq.~(\ref{vactheta})$. 

With the choice of the Grassmann odd ''basis vectors'' in the internal space of 
''Grassmann fermions'' and by extending these  ''basis states'' to momentum space 
to be able to solve the equations of motion, Eq.~(\ref{Weylgrass}),
we are able to define the creation operators 
$\underline{\hat {\bf b}}^{\theta k\,s }_{tot}(\vec{p})$ of the odd Grassmann character,
which together with their Hermitian conjugated partners annihilation operators, fulfill
the anticommutation relations of Eq.~(\ref{ijthetaprodgen}), manifesting the properties 
of the second quantized fermion fields. Anticommutation properties of creation and annihilation operators are due to the odd Grassmann character of the ''basis vectors''.

To define the Hilbert space of all possible ''Slater determinants'' of all possible 
occupied and empty fermion states and to discuss the application of 
$ \underline{\hat {\bf b}}^{\theta k\,s }_{tot} (\vec{p})$ and 
$ \underline{\hat {\bf b}}^{k\,s \dagger}_{tot} (\vec{p}) $ 
on ''Slater determinants'', let us see what the anticommutation relations, presented in Eq.~(\ref{ijthetaprodgen}), tell.
We realize from Eq.~(\ref{ijthetaprodgen}) the properties 
\begin{eqnarray}
\label{tensorproperties}
 \underline{\hat {\bf b}}^{\theta k\,s \dagger }_{tot} (\vec{p})*_{T} 
 \underline{\hat {\bf b}}^{\theta k'\,s' \dagger}_{tot} (\vec{p}{\,}')&=& - 
 \underline{\hat {\bf b}}^{\theta k'\,s' \dagger }_{tot} (\vec{p}{\,}')*_{T} 
 \underline{\hat {\bf b}}^{\theta k \,s \dagger}_{tot} (\vec{p})\,, \nonumber\\
  \underline{\hat {\bf b}}^{\theta k\,s  }_{tot} (\vec{p})*_{T} 
 \underline{\hat {\bf b}}^{\theta k'\,s' }_{tot} (\vec{p}{\,}')&=& - 
 \underline{\hat {\bf b}}^{\theta k'\,s'  }_{tot} (\vec{p}{\,}')*_{T} 
 \underline{\hat {\bf b}}^{\theta k \,s }_{tot} (\vec{p})\,, \nonumber\\
  \underline{\hat {\bf b}}^{\theta k\,s  }_{tot} (\vec{p})*_{T} 
 \underline{\hat {\bf b}}^{\theta k'\,s' \dagger}_{tot} (\vec{p}{\,}')&=& - 
 \underline{\hat {\bf b}}^{\theta k'\,s' \dagger }_{tot} (\vec{p}{\,}')*_{T} 
 \underline{\hat {\bf b}}^{\theta k \,s }_{tot} (\vec{p})\,, \nonumber\\
 {\rm if \;\, at \;\, least \,\; one \,\; of \,} (k,s, \vec{p}) && {\rm distinguishes\,\;from} 
(k',s', \vec{p}{\,}') \,,\nonumber\\
 \underline{\hat {\bf b}}^{\theta k\,s \dagger}_{tot} (\vec{p})*_{T} 
 \underline{\hat {\bf b}}^{\theta k\,s \dagger }_{tot} (\vec{p})
&=& 0\,,\nonumber\\ 
 \underline{\hat {\bf b}}^{\theta k\,s }_{tot} (\vec{p})*_{T} 
 \underline{\hat {\bf b}}^{\theta k\,s }_{tot} (\vec{p})
&=& 0\,,\nonumber\\ 
 \underline{\hat {\bf b}}^{\theta k\,s }_{tot} (\vec{p})*_{T}
 \underline{\hat {\bf b}}^{\theta k\,s \dagger}_{tot} (\vec{p}{\,}) |\,1>&=& |\,1>\,,\nonumber\\
 \underline{\hat {\bf b}}^{\theta k\,s }_{tot} (\vec{p}) |\,1>&=& 0\,,\nonumber\\
 |p^0|&=&|\vec{p}|\,.
\end{eqnarray}
From the above relations we recognize how do the creation and annihilation operators 
apply on ''Slater determinants'' of empty and occupied states, the later determined by 
 $ \underline{\hat {\bf b}}^{\theta k\,s \dagger }_{tot} (\vec{p})$:\\
 {\bf i.} The creation operator 
 $ \underline{\hat {\bf b}}^{\theta k\,s \dagger }_{tot} (\vec{p})$
jumps over the creation operator defining the occupied state, which distinguish from
the jumping creation one in at least one of ($k, s, \vec{p}$), changing sign of the
''Slater determinant'' every time, up to the last step when it comes to its own empty
state, the one with its quantum numbers ($k,s,\vec{p}$), 
occupying this empty state, or if this state is already occupied, gives zero. \\  
 {\bf ii.} The annihilation operator changes sign of the ''Slater determinant'' when 
ever jumping over the occupied state carrying different internal quantum numbers 
($k,s$) or  $\vec{p}$, unless it comes to the occupied state with its own ($k, s, \vec{p}$), emptying this state or, if this state is empty, gives zero.



We show in Part II that the Clifford odd ''basis vectors'' describe fermions with 
the half integer spin, offering as well the corresponding anticommutation relations,
explaining Dirac's postulates for second quantized fermions.

We discuss in Sect.~\ref{Hilbertspace0} the properties of the ''Slater determinants''
of the occupied and empty ''Grassmann fermion states'', created by  
$ \underline{\hat {\bf b}}^{\theta k\,s\dagger }_{tot} (\vec{p})$.

In Subsect.~B.a. we present one solution of the equations of motion for free
massless ''Grassmann fermions''.


\vspace{5mm}

$\;\;{\it B.a.}$ {\it Plane wave solutions of equations of motion}, Eq.~(\ref{Weylgrass}),
 {\it in $d=(5+1)$-dimensional space}

\vspace{2mm}

\begin{small}
One of such plane wave massless solutions of the equations of motion in 
$d=(5+1)$-dimensional space for momentum $p^a=(p^0, p^1, p^2, p^3, 0, 0)$, 
$p^0=|p^0|$, is the superposition of ''basis vectors'', presented in 
Table~\ref{Table grassdecuplet.} among the first three members of the first 
decuplet, $k=I$. This particular solution 
$ \underline{\hat{\bf b}}^{\theta k\, s \dagger}_{tot}\,(\vec{p})$ 
carries the spin ${\cal S}^{12}=1$ ("up") and the ``charge'' 
${\cal S}^{56}=1$ (both from the point of view of $d=(3+1)$) 
\begin{eqnarray}
\label{5+1sol}
&&  \underline{\hat{\bf b}}^{\theta 1 \,1 \dagger}_{tot}\,(\vec{p}) {\bf :} = 
\beta\, (\frac{1}{\sqrt{2}})^2 \{\,
\frac{1}{\sqrt{2}}\, (\theta^0 - \theta^3)  (\theta^1 +i \theta^2)
  \nonumber\\
&&  - \frac{2(|p^0| - |p^3|)}{p^1- i p^2} \,
  (\theta^0  \theta^3 +i \theta^1 \theta^2) \, \nonumber\\
&&- (\frac{(p^1 +i p^2)^2}{(|p^0| + |p^3|)^2}) \, \frac{1}{\sqrt{2}}
 (\theta^0 + \theta^3)  (\theta^1 - i \theta^2) \,\} \nonumber\\
&& \times  (\theta^5 +i \theta^6)\,
\cdot   e^{-i (|p^0| x^0 - \vec{p}\cdot \vec{x})}\,,
\;\; \quad |p^0| = |\vec{p}|\,,\nonumber
 \end{eqnarray}
%
 $\beta$ is the normalization factor. 
 The notation $ \underline {\hat{\bf b}}^{\theta 1\,1 \dagger}_{tot}\,(\vec{p})$  
 means that the creation operator represents the plane wave solution of the 
equations of motion, Eq.~(\ref{Weylgrass}), for a particular $|p^0|=|\vec{p}|$.

Applied on the vacuum state  the creation operator defines the second quantized 
single particle state of particular momentum $\vec{p}$. States, carrying different 
$\vec{p}$, are orthogonal (due to the orthogonality of the plane waves  of different 
momenta and  due to the orthogonality of 
$  \underline{\hat{\bf b}}^{\theta k' \,s' \dagger}_{tot}\,(\vec{p})$ and
$  \underline{\hat{\bf b}}^{\theta k \,s }_{tot}\,(\vec{p})$ with respect to $k$ and $s$, Eqs.~(\ref{ortpp'}, \ref{ortpp'con}, \ref{ijthetaprodgen})).

More solutions can be 
found in~\cite{nh2018} and the references therein.
\end{small}

\vspace{2mm}



%
\section{Hilbert space  of anticommuting integer spin ``Grassmann fermions''}
\label{Hilbertspace0}

The  Grassmann odd creation operators 
$  \underline{\hat {\bf b}}^{\theta\, k\,s \dagger}_{tot} (\vec{p})$, with
$|p^0|=|\vec{p}|$, are defined on the tensor products  of $2^{d-1}$ ''basis vectors'',
defining the internal space of integer spin ''Grassmann fermions'', and on infinite 
basis states of  momentum space for each component of $\vec{p}$, chosen so that 
they solve for particular $(\vec{p })$ the equations of motion, Eq.~(\ref{Weylgrass}). 
They  fulfill together with their Hermitian conjugated annihilation operators 
$  \underline{\hat {\bf b}}^{\theta\, k\,s}_{tot} (\vec{p})$ the
 anticommutation relations of Eq.~(\ref{ijthetaprodgen}).
 
These creation operators form the Hilbert space of ''Slater determinants'', defining for 
each ''Slater determinant'' places with either empty or occupied ''Grassmann fermion''
states. 

{\bf  Statement 3}: Introducing the tensor product multiplication $*_{T}$ of any 
number of single ''Grassmann fermion'' states of all possible internal quantum 
numbers  and all possible momenta (that is of any number of 
$  \underline{\hat {\bf b}}^{\theta\, k\,s \dagger}_{tot} (\vec{p})$ and with the identity included, applying on the vacuum state of any  $(k, s, \vec{p})$), we generate the Hilbert 
space of the second quantized ''Grassmann fermion'' fields.

It is straightforward to recognize that the above definition of the Hilbert space is equivalent 
to the space of ''Slater determinants'' of all possible empty or occupied states of any 
momentum and any quantum numbers  describing the internal space. The identity in 
this tensor product  multiplication, for example, represents the ''Slater determinant'' 
of no single fermion state present.

The $2^{d-1}$ Grassmann odd creation operators of particular momentum 
$\vec{p }$, if applied on the vacuum state $|\,1>$, Eq.~(\ref{vactheta}), 
define $2^{d-1}$ states. Since any state can be occupied or empty,
the Hilbert space  ${\cal H}_{\vec{p}}$ of a particular momentum $\vec{p}$
 consists correspondingly of 
\begin{eqnarray}
N_{{\cal H}_{\vec{p}}}& =&  2^{2^{d-1}}\,.
\label{NHpgrass}
\end{eqnarray}
''Slater determinants'', namely the one with no occupied state, those with one 
occupied state, those with two occupied states, up to the one with all $2^{d-1}$ states occupied. 

The total Hilbert space  ${\cal H}$ of anticommuting integer spin ''Grassmann fermions''
consists of infinite many ''Slater determinants'' of particular $\vec{p}$,  
${\cal H}_{\vec{p}}$, due to infinite many degrees of freedom in the momentum space
%
\begin{eqnarray}
{\cal H}& =& \prod_{\vec{p}}^{\infty}\otimes_{N} {\cal H}_{\vec{p}}\,,
\label{Hgrass}
\end{eqnarray}
with the infinite number of degrees of freedom
\begin{eqnarray}
N_{\cal H}& =& \prod_{\vec{p}}^{\infty} 2^{2^{d-1}}\,.
\label{NHgrass}
\end{eqnarray}

\subsection{''Slater determinants'' of anticommuting integer spin ``Grassmann fermions'' of particular momentum $\vec{p}$}
\label{Hilbertspacep}

Let us write down explicitly these $2^{2^{d-1}}$ contributions to the Hilbert space 
${\cal H}_{\vec{p}}$ of particular momentum ${\vec{p}}$, using the notation that 
${\bf 0^{k}_{s \vec{p}}}$ represents the unoccupied state 
$ \underline{\hat {\bf b}}^{\theta k\,s \dagger}_{tot} (\vec{p}) |\,1>$ (of the $s^{th}$ solution of the equations of motion belonging to the $k^{th}$ irreducible representation), while ${\bf 1^{k}_{s \vec{p}}}$ 
represents the corresponding occupied state.

 The number operator is according to Eq.~(\ref{ijthetaprod}) and Eq.~(\ref{tensorproperties}) equal to
 \begin{eqnarray}
\label{NOSDgrass}
&& N^{\theta k\, s}_{\vec{p}}=  
 \underline{\hat {\bf b}}^{\theta\,k\,s \dagger}_{tot} (\vec{p}) \,*_{T} 
  \underline{\hat {\bf b}}^{\theta \,k\,s }_{tot} (\vec{p}) \,,\nonumber\\
&& N^{\theta k s}_{ \vec{p}}\,*_{T}\, 0^{k}_{s\vec{p}}=0\,,\quad  
 N^{\theta k s}_{ \vec{p}} \, *_{T}\,1^{k}_{s\vec{p}}=1\,.
\end{eqnarray}

Let us simplify the notation so that we count for   $k=1$  empty states as 
${\bf 0_{r \vec{p}}}$, and occupied states as ${\bf 1_{r \vec{p}}}$, with 
$r=(1,\dots,  s^1_{max})$, 
for $k=2$ we count $r=s^1_{max}+1, \dots, s^1_{max}+ s^2_{max}$, 
up to $r= 2^{d-1}$.
Correspondingly we can represent  ${\cal H}_{\vec{p}}$ as follows
\begin{eqnarray}
\label{SDgrass}
|{\bf 0_{1\vec{p}}}, {\bf 0_{2 \vec{p}}}, {\bf 0_{3 \vec{p}}}, \dots,
{\bf  0_{2^{d-1} \vec{p}}}>\;\, ,&&\quad
|{\bf 1_{1\vec{p}}}, {\bf 0_{2 \vec{p}}}, {\bf 0_{3 \vec{p}}}, \dots,
{\bf  0_{2^{d-1} \vec{p}}}>,\nonumber\\
|{\bf 0_{1\vec{p}}}, {\bf 1_{2 \vec{p}}}, {\bf 0_{3 \vec{p}}}, \dots,
{\bf  0_{2^{d-1} \vec{p}}}>\;\,,&&\quad
|{\bf 0_{1\vec{p}}}, {\bf 0_{2 \vec{p}}}, {\bf 1_{3 \vec{p}}}, \dots,
{\bf  0_{2^{d-1} \vec{p}}}>,\nonumber\\
\vdots \nonumber\\
|{\bf 1_{1\vec{p}}}, {\bf 1_{2 \vec{p}}}, {\bf 0_{3 \vec{p}}}, \dots,
{\bf  0_{2^{d-1} \vec{p}}}>\;\,,&&\quad
|{\bf 1_{1\vec{p}}}, {\bf 0_{2 \vec{p}}},{\bf 1_{3 \vec{p}}}, \dots,
{\bf 0_{2^{d-1}\vec{p}}}>,\nonumber\\
\vdots\nonumber\\
|{\bf 1_{1\vec{p}}}, {\bf 1_{2 \vec{p}}}, {\bf 1_{3 \vec{p}}}, \dots,
{\bf  1_{2^{d-1} \vec{p}}}>\;\,,&&
\end{eqnarray}
with a part with none of states occupied ($N_{r\vec{p}}=0$ for all $r=1,\dots, 2^{d-1}$),
with a part with only one of states occupied ($N_{r \vec{p}}=1$ for a particular 
$r=1,\dots, 2^{d-1}$ while  $N_{r' \vec{p}}=0$ for all the others  $r' \ne r$),
with a part with only two of states occupied ($N_{r\vec{p}}=1$ and $N_{r'\vec{p}}=1$, where $r$ and $r'$ run from $1,\dots, 2^{d-1}$), and so on. The last part
has all the states occupied.

Taking into account Eq.~(\ref{tensorproperties}) is not difficult to see that the 
creation operator $ \underline{\hat {\bf b}}^{\theta k\,s \dagger}_{tot} (\vec{p})$ and the 
annihilation operators $ \underline{\hat {\bf b}}^{\theta k\,s }_{tot} (\vec{p})$, when applied on this Hilbert space ${\cal H}_{\vec{p}}$, fulfill the anticommutation relations for the second quantized ``fermions''.
\begin{eqnarray}
\{  \underline{\hat {\bf b}}^{\theta k\,s }_{tot} (\vec{p})\,,  
 \underline{\hat {\bf b}}^{\theta k'\,s' \dagger}_{tot} \, (\vec{p}{\,}) \}_{*_{T}+} 
{\cal H}_{\vec{p}} &=& \delta^{k k'}\; \delta_{s s'} 
{\cal H}_{\vec{p}}\,,\nonumber\\
\{   \underline{\hat {\bf b}}^{\theta k\,s }_{tot} (\vec{p})\,,  
 \underline{\hat {\bf b}}^{\theta k'\,s'}_{tot} (\vec{p}{\,}) \}_{*_{T}+}\;  
{\cal H}_{\vec{p}}&=& 0\;\cdot \,{\cal H}_{\vec{p}} \,,\nonumber\\
\{  \underline{\hat {\bf b}}^{\theta k\,s \dagger}_{tot} (\vec{p})\, ,
  \underline{\hat {\bf b}}^{\theta k'\,s' \dagger}_{tot} \,(\vec{p}{\,}) \}_{*_{T}+}\;
  {\cal H}_{\vec{p}}&=&
0\;\cdot\,{\cal H}_{\vec{p}}\,.
\label{ijthetaprodgenHpT}
\end{eqnarray}

The proof for the above relations easily follows if taking into account that, when ever 
the creation or annihilation operator jumps over an odd products of occupied states, 
the sign changes.
Then one sees that the contribution of the application of 
$ \underline{\hat {\bf b}}^{\theta k\,s }_{tot} (\vec{p})*_{T}$
$  \underline{\hat {\bf b}}^{\theta k'\,s' \dagger}_{tot} (\vec{p})\; {\cal H}_{\vec{p}}$ has
the opposite sign than the contribution of  
$  \underline{\hat {\bf b}}^{\theta k'\,s' \dagger}_{tot} (\vec{p})*_{T}$
$ \underline{\hat {\bf b}}^{\theta k\,s}_{tot} (\vec{p})\; {\cal H}_{\vec{p}}$. 

If the creation and annihilation operators are Hermitian conjugated to each other, 
the result of 
\[\{  \underline{\hat {\bf b}}^{\theta k\,s }_{tot} (\vec{p})\,*_{T}
  \underline{\hat {\bf b}}^{\theta k\,s \dagger}_{tot} (\vec{p}) +
  \underline{\hat {\bf b}}^{\theta k\,s \dagger}_{tot} (\vec{p}) \,*_{T}
  \underline{\hat {\bf b}}^{\theta k\,s }_{tot} (\vec{p})\,\} {\cal H}_{\vec{p}}= 
 {\cal H}_{\vec{p}}\]
 is the whole ${\cal H}_{\vec{p}}$ back. Each of the two summands operates on its
  own half of ${\cal H}_{\vec{p}}$. Jumping together over even number of occupied states
 $  \underline{\hat {\bf b}}^{\theta k\,s }_{tot} (\vec{p})$ and 
 $  \underline{\hat {\bf b}}^{\theta k\,s \dagger}_{tot} (\vec{p})$ do  not change the sign of particular ``Slater determinant''.  
(Let us add that  $  \underline{\hat {\bf b}}^{\theta k\,s }_{tot} (\vec{p})$  reduces for particular $k$ and $s$ the Hilbert space ${\cal H}_{\vec{p}}$ for a factor  $\frac{1}{2}$, and so does 
$  \underline{\hat {\bf b}}^{\theta k\,s \dagger}_{tot} (\vec{p})$. 
The sum of both, applied on ${\cal H}_{\vec{p}}$, reproduces the whole 
${\cal H}_{\vec{p}}$.)

\subsection{''Slater determinants'' of Hilbert space  of anticommuting integer spin ``fermions'' }
\label{Hilbertspacegen}

The total Hilbert space of  anticommuting ''fermions'' is the infinite product  
of the Hilbert spaces of particular $\vec{p}$, 
${\cal H} = \prod_{\vec{p}}^{\infty}\otimes_{N} {\cal H}_{\vec{p}}$, Eq.~(\ref{Hgrass}), represented by infinite numbers of ''Slater determinants'' $N_{\cal H} = \prod_{\vec{p}}^{\infty} 2^{2^{d-1}}$, Eq.~(\ref{NHgrass}).
%
%
%
The notation $\otimes_{N}$ is to point out that  creation operators 
$ \underline{\hat {\bf b}}^{\theta k\,s \dagger}_{tot} (\vec{p}{\,})$, which origin in
superposition of odd number of $\theta^a$'s, keep their odd  character also 
in the tensor products of the internal and momentum space, as well as in the ''Slater 
determinants'', in which creation operators determine the occupied states. 

The application of creation operators
$ \underline{\hat {\bf b}}^{\theta k\,s \dagger}_{tot} (\vec{p}{\,})$ and their Hermitian conjugated annihilation operators 
$ \underline{\hat {\bf b}}^{\theta k\,s}_{tot} (\vec{p}{\,})$ on the Hilbert space 
${\cal H}$ has the property, manifested in Eq.~(\ref{tensorproperties}), leading to the conclusion that the application of 
$ \underline{\hat {\bf b}}^{\theta k\,s \dagger}_{tot} (\vec{p}{\,})*_{T}$
$ \underline{\hat {\bf b}}^{\theta k'\,s' \dagger}_{tot} (\vec{p'}{\,}) *_{T} {\cal H}$ is
not zero if at least one of $(k, s, \vec{p})$ is not equal to $(k', s', \vec{p'})$,
while $ \underline{\hat {\bf b}}^{\theta k\,s \dagger}_{tot} (\vec{p}{\,})*_{T}$
$ \underline{\hat {\bf b}}^{\theta k'\,s' \dagger}_{tot} (\vec{p'}{\,}) *_{T} {\cal H}+$
$ \underline{\hat {\bf b}}^{\theta k'\,s' \dagger}_{tot} (\vec{p'}{\,})*_{T}$
$ \underline{\hat {\bf b}}^{\theta k\,s \dagger}_{tot} (\vec{p}{\,}) *_{T} {\cal H}=0$
for any $(k, s, \vec{p})$ and any $(k', s', \vec{p'})$, what is not difficult to prove when
taking into account Eq.~(\ref{tensorproperties}).

%
%
One can easily show that the creation operators $  \underline{\hat {\bf b}}^{\theta k\,s \dagger}_{tot} (\vec{p})$ and the annihilation operators 
$  \underline{\hat {\bf b}}^{\theta k\,s }_{tot} (\vec{p}{\,}')$ fulfill equivalent anticommutation  on the whole Hilbert space of infinity many ''Slater determinants'' as 
they do on the Hilbert space ${\cal H}_{\vec{p}}$.
\begin{eqnarray}
\{   \underline{\hat {\bf b}}^{\theta k\,s }_{tot} (\vec{p})\,,  
 \underline{\hat {\bf b}}^{\theta k\,s \dagger}_{tot} (\vec{p}{\,}') \}_{*_{T}+} 
{\cal H} &=& \delta^{k k'}\; \delta_{s s'} \delta (\vec{p} -\vec{p}{\,}')\;
{\cal H}\,,\nonumber\\
\{  \underline{\hat {\bf b}}^{\theta k\,s}_{tot} (\vec{p}),  
 \underline{\hat {\bf b}}^{\theta k\,s \dagger}_{tot} (\vec{p}{\,}') \}_{*_{T}+}\;  
{\cal H}&=& 0\;\cdot\,{\cal H} \,,\nonumber\\
\{  \underline{\hat {\bf b}}^{\theta k\,s \dagger}_{tot} (\vec{p})\, ,
  \underline{\hat {\bf b}}^{\theta k'\,s' \dagger}_{tot} (\vec{p}{\,}') \}_{*_{T}+}\; {\cal H}&=& 0\;\cdot\,{\cal H}\,.
\label{ijthetaprodgenHT}
\end{eqnarray}
%


Creation operators, $ \underline{\hat{\bf b}}_{tot}^{s f \dagger} (\vec{p})$, operating on a vacuum state, as well as on the whole Hilbert space, define the second quantized fermion 
states.

\subsection{Relations between  creation operators 
$ \underline{\hat {\bf b}}^{\theta \,k\,s \dagger}_{tot} (\vec{p})$ in the Grassmann odd algebra and the creation operators postulated by Dirac }
\label{comparisonND}

Creation operators $ \underline{\hat{\bf b}}_{tot}^{\theta \,k\,s \dagger} (\vec{p})$ define 
the second quantized ''fermion''  fields of integer spins. 

Since the second quantized Dirac fermions have the half integer spin, the ''Grassmann 
fermions'', the internal degrees of which is described by the Grassmann odd algebra, 
have the integer spin. The comparison between the second quantized fields of Dirac and 
those presented in this Part I of the paper can only be done on a rather general level.
We leave therefore the detailed comparison of the creation and 
annihilation operators for fermions with half integer spins between those postulated 
by Dirac and the  ones following from the Clifford odd algebra presented in Part II to Subsect.~3.4 of Part II.

Here we discuss only the relations among appearance of the creation and annihilation 
operators offered by the Grassmann odd algebra and those postulated by Dirac. In 
both cases we treat only $d=(3+1)$-dimensional space, that is we solve the equations of motion for $p^a=(p^0, p^1, p^2, p^3)$ (in the case that $d >4$ 
the rest of space demonstrates the charges in $d=(3+1)$, when 
$p^a=(p^0, p^1, p^2, p^3, 0, 0, \dots, 0)$).


 It is pointed out in what follows that both internal spaces --- either the internal 
space postulated by Dirac or the internal space offered by the Grassmann algebra --- 
are finite dimensional, as also the internal space offered by the Clifford algebra is finite dimensional. 

In the Dirac case the second quantized states  are in $d=(3+1)$ dimensions 
postulated as follows
\begin{eqnarray}
\label{stateDirac}
{\underline {\bf {\Huge \Psi}}}^{s \dagger}(x^0,\vec{x}) & =&
\sum_{i, \vec{p}_k} \,\hat{{\bf a}}^{\dagger}_i (\vec{p}_k)\, 
u^s_i (\vec{p}_k)\,e^{-i(p^0 x^0 -\varepsilon \vec{p}\cdot \vec{x})}\,. 
\end {eqnarray}
$v^s_i (\vec{p}_k)\, (= u^s_i \, e^{-i(p^0x^0)-\varepsilon \vec{p}\cdot \vec{x}})$ 
are the two  left handed  ($\Gamma^{(3+1)}=-1$) and the two right 
handed ($\Gamma^{(3+1)}= 1$, Eq.~(B.3)) two-component column matrices, 
representing the four solutions $s$ of the Weyl equation for free massless fermions 
of particular momentum $|\vec{p}_k|= |p_{k}^{0}|$~(\cite{BetheJackiw},
 Eqs.~(20-49) - (20-51)), the factor $\varepsilon =\pm1$ depends on the product of 
 handedness and spin.

${\hat{\bf a}}^{\dagger}_i (\vec{p}_k)$ are by Dirac postulated creation operators, 
which together  with annihilation operators $\hat{{\bf a}}_i (\vec{p}_k)$, fulfill the 
anticommutation relations~(\cite{BetheJackiw}, Eqs.~(20-49) - (20-51)),
 \begin{eqnarray}
 \label{comDirac}
\{{\hat{\bf a}}^{\dagger}_i (\vec{p}_k), \,{\hat{\bf a}}^{\dagger}_j (\vec{p}_l)\}_{*_{T}+}&=&
0= \{{\hat{\bf a}}_i (\vec{p}_k), \,{\hat{\bf a}}_j (\vec{p}_l)\}_{*_{T}+}\,,\nonumber\\ 
\{{\hat{\bf a}}_i (\vec{p}_k), \,{\hat{\bf a}}^{\dagger}_j (\vec{p}_l)\}_{*_{T}+} &=&
\delta_{ij}\delta_{\vec{p}_k \vec{p}_l}\,,
\end{eqnarray}
in the case of discretized  momenta for a fermion in a box.   Creation operators 
and annihilation operators, ${\hat{\bf a}}^{\dagger}_i (\vec{p}_k)$ and 
${\hat{\bf a}}_i (\vec{p}_k)$, are postulated to have on  the Hilbert space of all 
''Slater determinants'' these anticommutation properties. 


To be able to  relate the creation operators of Dirac 
$ {\hat{\bf a}}^{\dagger}_i (\vec{p}_k)$  with 
$ \underline{\hat{\bf b}}^{\theta k s \dagger}_{tot} (\vec{p}_k)$ from Eq.~(\ref{ijthetaprodgenHT}), let us remind the reader that 
$ \underline{\hat{\bf b}}^{\theta k s \dagger}_{tot} (\vec{p}_k)$ is a superposition of 
basic vectors $\hat{b}^{\theta k \dagger}_i$ with the coefficients 
$c^{ks}{}_{i}(\vec{p})$,  which depend on the momentum $\vec{p}$, Eq.~(\ref{ptheta}) 
(${\hat{\bf b}}^{\theta k \,s \dagger} (\vec{p}) = 
\sum_{i} c^{k s}{}_{i} (\vec{p}) \, \hat{b}^{\theta k \dagger}_{i} $),
so that $ \underline{\hat{\bf b}}^{\theta k s \dagger}_{tot} (\vec{p}_k)$
($=\sum_{i} c^{k s}{}_{i} (\vec{p}) \, \hat{b}^{\theta k \dagger}_{i}
\, e^{-i(p^0 x^0 -\varepsilon \vec{p}\cdot \vec{x})} $)  solves the
equations of motion for free massless ''Grassmann fermions'' for plane waves, while 
$|p^0|=|\vec{p}|$.  

We treat  in this subsection the Grassmann case in $(3+1)$-dimensional space only, without taking care on different irreducible representations $k$ as well as on charges, in order to be
able to relate the creation and annihilation operators in Grassmann space with the Dirac's ones. In this case the odd Grassmann creation operators are expressible with the ''basic vectors'', which are fourplets, presented in Table~\ref{Table grassdecuplet.} on the $7^{th}$ up to 
 the $10^{th}$ lines, the same on both decuplets, neglecting $\theta^5 \theta^6$ contribution. 
(They have handedness in $d=(3+1)$ equal zero.)

Let us rewrite creation operators in the Dirac case so that their expressions resemble the 
expression for the creation operators $ \underline{\hat{\bf b}}^{\theta  s \dagger}_{tot} (\vec{p}_k) =\sum_{i} c^{ s}{}_{i} (\vec{p}) \, \hat{b}^{\theta  \dagger}_{i}
\, e^{-i(p^0 x^0 -\varepsilon \vec{p}\cdot \vec{x})} $, leaving out the index of the
irreducible representation.
\begin{eqnarray}
\label{relDirac}
\underline{\bf {\hat a}}^{ s \dagger}_{tot} (\vec{p}_k) & \stackrel{\mathrm{def}}{=}& \sum_{i} \,\hat{{\bf a}}^{\dagger}_i (\vec{p}_k)\, 
u^s_i (\vec{p}_k)\,e^{-i(p^0 x^0 -\varepsilon \vec{p}\cdot \vec{x})}
\stackrel{\mathrm{def}}{=}
\sum_{i} \alpha^{s}_{i} (\vec{p}_{k}) \, {\hat a}^{\dagger}_{i} \,
e^{-i(p^0 x^0 -\varepsilon \vec{p}\cdot \vec{x})}\,\nonumber\\
&& {\rm to \;be \;compared \, with}\nonumber\\
\underline{\hat{\bf b}}^{\theta  s \dagger}_{tot} (\vec{p}_k) &=&\sum_{i} c^{ s}{}_{i} (\vec{p}) \, \hat{b}^{\theta  \dagger}_{i}
\, e^{-i(p^0 x^0 -\varepsilon \vec{p}\cdot \vec{x})}\,.
\end{eqnarray}
%

%

 %
We define in the Dirac case two  creation operators: $\underline{\bf {\hat a}}^{ s \dagger}_{tot} (\vec{p}_k)$ and $\hat{a}^{\dagger}_i$. 
 Since ${\underline {\bf {\Huge \Psi}}}^{s \dagger}(x^0,\vec{x}) =\sum_{\vec{p}_k}\, 
\underline{\bf {\hat a}}^{ s \dagger}_{tot} (\vec{p}_k)$, Eq.~(\ref{stateDirac}),
we realize that the two expressions $u^s_i (\vec{p}_k)\, 
\hat{{\bf a}}^{\dagger}_i (\vec{p}_k)$ and 
$ \alpha^{s}_{i} (\vec{p}_{k}) \, {\hat a}^{\dagger}_{i}$
describe the same degrees of freedom.

These new creation operators $\underline{\bf {\hat a}}^{ s \dagger}_{tot} (\vec{p}_{k})$
can not be related directly to 
$\underline{\bf {\hat b}}^{\theta  s \dagger}_{tot} (\vec{p}_{k})$, since the first ones describe the second quantized fields of the half integer spin fermions, while the later describe 
the second quantized integer spin ''fermion'' fields. However, both fulfill the anticomutation relations of Eq.~(\ref{ijthetaprodgenHT}).

The reader can notice that the creation operators ${\hat a}^{\dagger}_{i}$ do not 
depend on $\vec{p}$ as also ${\hat b}^{\theta \dagger}_{i}$ do not, both describing 
the internal degrees of freedom, while $\alpha^{s}_{i} (\vec{p}_{k}) \, 
{\hat a}^{\dagger}_{i}$  and $\alpha^{s}_{i} (\vec{p}_{k}) \, 
{\hat b}^{\theta \dagger}_{i}$ do.


The creation and annihilation operators of Dirac fulfill obviously the anticommutation 
relations of Eq.~(\ref{ijthetaprodgenHT}). To see this we only have to replace 
$\underline{\hat{\bf b}}^{\theta h\,s \dagger}_{tot} (\vec{p})$ by 
$\underline{\hat{\bf a}}^{ h\,s \dagger}_{tot} (\vec{p})$ by taking into account relation
of Eq.~(\ref{relDirac}).

Creation and annihillation operators of the Dirac second quantized fermions with half integer spins are  in Part II, in Subsect.~III.D, related to the corresponding ones, offered by the 
Clifford algebra. Relating the creation and annihilation operators offered by the Clifford 
algebra  objects with the Dirac's ones ensures us that the Clifford odd algebra explains the Dirac's postulates.   


%
\section{Conclusions}
\label{conclusionsGrass}

We learn in this Part I paper, that  in $d$-dimensional space the superposition of odd 
products of  $\theta^a$'s exist, 
Eqs.~(\ref{start(2n+1)2theta}, \ref{start(2n+1)2thetasecond}, \ref{start4ntheta}), chosen to be the eigenvectors of the Cartan subalgebra, Eq.~(\ref{eigengrasscartan}),  
which together with their Hermitian conjugated partners, odd products of $\frac{\partial}{\partial \theta_a}$'s, Eqs.~(\ref{thetaderher}, \ref{start(2n+1)2theta}, \ref{grasscartanher}), fulfill  on the algebraic level 
on the vacuum state $|\phi_o>=|\,1>$, Eq.~(\ref{ijthetaprodgen}),
the requirements for the anticommutation relations for the Dirac's fermions. 

The creation operators defined on the tensor products of internal space of ''Grassmann 
basis vectors'' (of finite number of basis states) and of momentum space (with infinite 
number of basis states),  arranged to be solutions of the equation of motion for free massless ''Grassmann fermions'',  Eq.~(\ref{Weylgrass}),
 form the infinite dimensional  Hilbert space of  ''Slater determinants'' of  (continuous) infinite number of momenta, with
 $2^{2^{d-1}}$ possibilities for each momentum $\vec{p}$, Eq.~(\ref{ijthetaprodgenHT})).
These creation operators and their Hermitian conjugated partners fulfill on the Hilbert space 
the anticommutation relations postulated by Dirac for second quantized fermion fields.

 We demonstrate the way of deriving second quantized integer fermion fields. 
 
 In the subsection~\ref{HNsubsection} we clarify the relation between our description
 of the internal space of fermions with ''basis vectors'', manifesting oddness and transferring
 the oddness to the corresponding creation and annihilation operators of second quantized
 fermions, to the ordinary second quantized creation and annihilation operators from a 
 slightly different  point of view.

Since the  creation and annihilation operators, which are superposition of odd products of 
$\theta^a$'s and $\frac{\partial}{\partial \theta_a}$'s, respectively, anticommute 
algebraically when applying on the vacuum state, Eq.~(\ref{ijthetaprod}, \ref{ijthetaphi}) 
(while the  corresponding even products of $\theta^a$'s and 
$\frac{\partial}{\partial \theta_a}$'s commute, 
Eq.~(\ref{ijthetaprodeven})), it follows that also creation operators, defined on 
tensor products of the finite number of ''basis vectors'' (describing the internal degrees of freedom of ''Grassmann fermions'') and on infinite basis of 
momentum space, together with their Hermitian conjugated partners annihilation operators, fulfill the anticommutation relations of Eq.~(\ref{ijthetaprodgenHT}). 
The use of the Grassmann odd algebra to describe the internal space of 
''Grassmann fermions'' offers the anticommutation relations without postulating them:
on the (simple) vacuum state as well as on the Hilbert space of infinite number of ''Slater
determinants'' of all possible single particle states, empty or occupied, of the second 
quantized integer spin ''fermion'' fields. Correspondingly we second quantized ''fermion fields''
without postulating commutation relations of Dirac.

The internal ''basis vectors'' are chosen to be eigenvectors of the Cartan subalgebra 
operators in the way that the symmetry agrees with the properties of usual Dirac's 
creation and annihilation operators of second quantized fermions --- in the Clifford case
for half integer spin, while in the ''Grassmann fermions'' for the integer spins. 

 The ''Grassmann fermions" carry the spin and charges, originated in 
 $d\ge5$, in  the adjoint representations.  
''Grassmann fermions" offer no families, what means that there is no  available  
operators, which would connect different irreducible representations of the Lorentz group 
(without breaking symmetries). 


No elementary "Grassmann fermions" with the spins and charges in the adjoint 
representations have been observed, and since the observed quarks and leptons 
and anti-quarks and anti-leptons have half integer spins, charges in the
fundamental representations and appear in families, it does not seem possible for 
the future observation of the integer spin elementary "Grassmann fermions", especially not 
since Eq.~(19) in Part II demonstrates that the reduction of space in Clifford case, 
needed for the appearance of second quantized half integer  fermions, reduces also  the Grassmann space, leaving no place for second quantized ''Grassmann fermions'' with 
the integer spin.
 
In Part II two kinds of operators are studied; There are namely two kinds of the Clifford algebra objects, $\gamma^a= (\theta^{a} + \frac{\partial}{\partial \theta_{a}})$ and $\tilde{\gamma}^{a} =i \,(\theta^{a} - \frac{\partial}{\partial \theta_{a}})$, which anticommute, $\{\gamma^a, \tilde{\gamma}^a \}_{+} =0$ 
($\{\gamma^a, \gamma^b \}_{+} = 2 \eta^{ab}=$ 
$\{\tilde{\gamma}^a, \tilde{\gamma}^b \}_{+}$),
and offer correspondingly two kinds of independent representations. 

Each of these two kinds of independent representations can be arranged into irreducible representations with respect to the two Lorentz generators --- 
$S^{ab}= \frac{i}{4}\, (\gamma^{a} \gamma^{b} - \gamma^{b} \gamma^{a})$ and $\tilde{S}^{ab}= \frac{i}{4}\, (\tilde{\gamma}^{a}\tilde{\gamma}^{b} - \tilde{\gamma}^{b}\tilde{\gamma}^{a})$. 
All the Clifford irreducible representations of any of the two kinds of algebras are 
independent and disconnected. 

The two Dirac's actions in $d$-dimensional space for free massless fermions  
(${\cal A}\,  = \int \; d^dx \; \frac{1}{2}\, (\psi^{\dagger}\gamma^0 \, 
\gamma^a \,p_{a} \psi) +  h.c. $  and 
$\tilde{\cal{A}}\,  = \int \; d^dx \; \frac{1}{2}\, 
(\psi^{\dagger}\tilde{\gamma}^0 \, \tilde{\gamma}^a \, p_{a} \psi) + h.c. $ ) lead to the equations of motion, which have the solutions in both kinds of algebras for an odd Clifford character (they are superposition of an odd products of $\gamma^{a}$'s and $\tilde{\gamma}^a$'s, respectively), forming on the tensor product of finite number of ''basis vectors''  
describing the internal space and of the infinite number of basis of momentum space, the creation and annihilation operators, which only "almost" anticommute, while the 
Grassmann odd creation and annihilation operators do anticommute.
Although "vectors" of one irreducible representation of an odd Clifford algebra character, anticommute among themselves and so do  their Hermitian conjugated partners in each of the two kinds of the Clifford algebras, the anticommutation relations  among creation and annihilation operators in each of the two Clifford algebras separately, do not fulfill the  requirement, that only the anticommutator 
of a creation operator and its Hermitian conjugated partner gives a nonzero contribution. 

The decision, the postulate, Eq.~(12), that only one kind of the Clifford algebra objects --- 
we make a choice of $\gamma^a$ --- describes the internal space of fermions, 
while the second kind --- $\tilde{\gamma}^{a}$ in this case ---  does  not, 
and  consequently  determine ``family'' quantum numbers which distinguish 
among irreducible representations of  $S^{ab}$, solves the problems: \\
{\bf a.} Creation operators and their Hermitian conjugated partners, which are 
odd products of superpositions of $\gamma^a$, applied on the vacuum state,
 fulfill on the algebraic level the anticommutation relations, and the creation and annihilation operators creating the second quantized Clifford fermion fields fulfill all the requirements, 
which Dirac postulated for fermions. \\
{\bf b.} Different irreducible representations with respect to $S^{ab}$ carry 
now different "family" quantum numbers determined by $\frac{d}{2}$ 
commuting operators among $\tilde{S}^{ab}$.\\   
{\bf c.} The operators of the Lorentz algebra, 
which do not belong to the Cartan subalgebra, 
connect different irreducible representations of $S^{ab}$.

The above mentioned decision, Eq.~(19) in Part II, obviously reduces the degrees 
of freedom of the odd 
(and even) Clifford algebra, while opening the possibility for the appearance of 
''families'',  as well as for the explanation for the Dirac's second quantization postulates. 
This decision, reducing as well the degrees of freedom  of Grassmann algebra, 
disables the existence of the integer spin  "fermions" 
as elementary particles.  

Let us point out again at the end that when the internal part of the single particle 
wave function anticommute under the algebra product $*_A$,  then this
implies that the wave functions with such internal part
anticommute under the extension of $*_A$ to the (full) single particle wave
functions and so do anticommute the corresponding  creation and annihilation 
operators what manifests also on the properties of the Hilbert space. 

The anticommuting single fermion states manifest correspondingly the oddness 
already on the level of the first quantization.


\appendix
\section{Norms in Grassmann space and Clifford space}
\label{normgrass}
%


Let us define the integral over the Grassmann  space~\cite{norma93} of two functions of the 
Grassmann coordinates $<{\cal {\bf B}}|\theta> <{\cal {\bf C}}|\theta>$, $<{\cal {\bf B}}| \theta>= 
<\theta | {\cal {\bf B}}>^{\dagger}$,
\[<{\cal {\bf b}}|\theta>= \sum_{k=0}^{d} b_{a_1\dots a_k}
\theta^{a_1}\cdots \theta^{a_k},\]
 by requiring 
\begin{eqnarray}
\label{grassintegral}
&&\{ d\theta^a, \theta^b \}_{+} =0\,, \,\;\;  \int d\theta^a  =0\,,\,\;\; 
\int d\theta^a \theta^a =1\,,\;\; \nonumber\\
&&\int d^d \theta \,\,\theta^0 \theta^1 \cdots \theta^d =1\,,
\nonumber\\
&&d^d \theta =d \theta^d \dots d\theta^0\,,\,\;\; 
\omega = \prod^{d}_{k=0}(\frac{\partial}{\;\,\partial \theta_k} + \theta^{k})\,,
\end{eqnarray}
with $ \frac{\partial}{\;\,\partial \theta_a} \theta^c = \eta^{ac}$. We shall use the weight function~%
\cite{norma93} 
$\omega= \prod^{d}_{k=0}(\frac{\partial}{\;\,\partial \theta_k} + \theta^{k})$ to define the scalar
product  in Grassmann space $<{\cal {\bf B}}|{\cal {\bf C}} >$ 
\begin{eqnarray}
\label{grassnorm}
<{\cal {\bf B}}|{\cal {\bf C}} > &=&  \int 
 d^d \theta^a\, \,\omega 
 <{\cal {\bf B}}|\theta>\, <\theta|{\cal {\bf C}}> \nonumber\\
 &=& \sum^{d}_{k=0} \int 
\, b^{*}_{b_{1} \dots b_{k}} c_{b_1 \dots b_{k}}\,.%
\end{eqnarray}%
%

 To define norms in Clifford space Eq.~(\ref{grassintegral}) can be used as well.


%
\section{Handedness in Grassmann and Clifford space}
\label{handednessGrassCliff}

The handedness $\Gamma^{(d)}$ is one of the invariants of the group $SO(d)$, 
with the infinitesimal generators of the Lorentz group $S^{ab}$,
defined as 
\begin{eqnarray}
\label{handedness}
\Gamma^{(d)}&=&\alpha \varepsilon_{a_1 a_2\dots a_{d-1}} a_d\, S^{a_1 a_2} 
\cdot S^{a_3 a_4} \cdots S^{a_{d-1} a_d}\,,
\end{eqnarray}
with $\alpha$, which is chosen so that $\Gamma^{(d)}=\pm 1$.

In the Grassmann case  $S^{ab}$  is defined in Eq.~(\ref{thetasab}), while in the Clifford case
Eq.~(\ref{handedness}) simplifies, if we take into account that $S^{ab}|_{a\ne b}= 
\frac{i}{2}\gamma^a \gamma^b$  and $\tilde{S}^{ab}|_{a\ne b}= 
\frac{i}{2}\tilde{\gamma}^a \tilde{\gamma}^b$, as follows
\begin{eqnarray}
\Gamma^{(d)} :&=&(i)^{d/2}\; \;\;\;\;\;\prod_a \quad (\sqrt{\eta^{aa}} \gamma^a), 
\quad {\rm if } \quad d = 2n\,. 
\nonumber\\
\label{hand}
\end{eqnarray}

\begin{acknowledgements}
The author N.S.M.B. thanks Department of Physics, FMF, University of Ljubljana, Society of 
Mathematicians, Physicists and Astronomers of Slovenia,  for supporting the research on the 
{\it spin-charge-family} theory, the author H.B.N. thanks the Niels Bohr Institute for
being allowed to staying as emeritus, both authors thank DMFA and  Matja\v z Breskvar of 
Beyond Semiconductor for donations, in particular for sponsoring the annual workshops entitled 
"What comes beyond the standard models" at Bled. 
\end{acknowledgements}



\end{document}